\newcount\draft\draft=0  % Change to 0 for submission.
\documentclass[10pt, conference, compsocconf]{IEEEtran}
\usepackage{cite}
\usepackage[cmex10]{amsmath}
\usepackage{amssymb,amsfonts}
\usepackage{algorithmic}
\usepackage{graphicx}
\usepackage{textcomp}
\usepackage{xcolor}
\usepackage[hyphens]{url}
\usepackage{listings}
\usepackage{subcaption}
\usepackage[capitalize]{cleveref}
\usepackage{multirow}
\usepackage{etoolbox}
\usepackage{tikz}
\usepackage{xxx}
\renewcommand{\figurename}{Figure}

\def\BibTeX{{\rm B\kern-.05em{\sc i\kern-.025em b}\kern-.08em
    T\kern-.1667em\lower.7ex\hbox{E}\kern-.125emX}}

\definecolor{light-gray}{gray}{0.95}
\lstdefinelanguage{mlir}{
  sensitive=false,
  keywords={if,else,for,to,par_for},
  comment=[l]{//},
}

\lstdefinelanguage{aiengine}{
  sensitive=false,
  keywords={mul4, mac4},
  comment=[l]{//},
}

\lstdefinestyle{figurestyle}{
  basicstyle=\footnotesize\ttfamily,
  numbers=left,
  numberstyle=\footnotesize\sffamily,
  numbersep=0.7em,
  stepnumber=1,
  firstnumber=1,
}
\lstdefinestyle{normalstyle}{
  basicstyle=\footnotesize\ttfamily,
  numbers=none
}
\newtoggle{infig}
\togglefalse{infig}
\AtBeginEnvironment{figure}{\toggletrue{infig}}
\AtEndEnvironment{figure}{\togglefalse{infig}}
\AtBeginEnvironment{figure*}{\toggletrue{infig}}
\AtEndEnvironment{figure*}{\togglefalse{infig}}

\AtBeginEnvironment{lstlisting}{
  \iftoggle{infig}{
    \lstset{style=figurestyle,numbers=none}
  }{
    \lstset{style=normalstyle,numbers=none}}}

\newcommand{\equeue}{EQueue dialect}
\newcommand{\code}{\lstinline[language=]}
\newcommand*\circled[1]{\tikz[baseline=(char.base)]{
            \node[shape=circle,draw,inner sep=2pt] (char) {#1};}}
\title{Compiler-Driven Simulation \\ of Reconfigurable Hardware Accelerators} 
\author{\IEEEauthorblockN{Zhijing Li\IEEEauthorrefmark{1}, 
Yuwei Ye\IEEEauthorrefmark{1},
Stephen Neuendorffer\IEEEauthorrefmark{2},
Adrian Sampson\IEEEauthorrefmark{1}}
\IEEEauthorblockA{
\IEEEauthorrefmark{1}Cornell University, 
\IEEEauthorrefmark{2}Xilinx Inc.
}
\IEEEauthorblockA{
\IEEEauthorrefmark{1}\{zl679, yy453, asampson\}@cornell.edu,
\IEEEauthorrefmark{2}stephenn@xilinx.com}
}

\begin{document}
\maketitle
\thispagestyle{plain}
\pagestyle{plain}

%%%%%% -- PAPER CONTENT STARTS-- %%%%%%%%

\begin{abstract}
As customized accelerator design has become increasingly popular to keep up with the demand for high performance computing, it poses challenges for modern simulator design to adapt to such a large variety of accelerators. Existing simulators tend to two extremes: low-level and general approaches, such as RTL simulation, that can model any hardware but require substantial effort and long execution times; and higher-level application-specific models that can be much faster and easier to use but require one-off engineering effort.

This work proposes a compiler-driven simulation workflow that can model configurable hardware accelerator. The key idea is to separate structure representation from simulation by developing an intermediate language that can flexibly represent a wide variety of hardware constructs. We design the Event Queue (EQueue) dialect of MLIR, a dialect that can model arbitrary hardware accelerators with explicit data movement and distributed event-based control; we also implement a generic simulation engine to model EQueue programs with hybrid MLIR dialects representing different abstraction levels. We demonstrate two case studies of EQueue-implemented accelerators: the systolic array of convolution and SIMD processors in a modern FPGA.
In the former we show EQueue simulation is as accurate as a state-of-the-art simulator, while offering higher extensibility and lower iteration cost via compiler passes.
In the latter we demonstrate our simulation flow can guide designer efficiently improve their design using visualizable simulation outputs.
%We demonstrate our simulation flow can help designer efficiently try out different designs and recognize the bottleneck and visualizable outputs.


\end{abstract}

\section{Introduction}
% \begin{itemize}
%     \item components, bandwidth, more systematic design 
%     \item simulators are usually specific to some back end design and the parameters are 
%     \item with a small change in the back end design, the simulator needs to be modified significantly.
%     \item allows multiple levels of abstraction, 
%     \item never changes the simulator, but for existing ones we need to change
%     \item no need to redesign simulator, because we can
%     \item scalesim input/
% \end{itemize}

% 1. exisiting simulators
% 2. how are they problematic. and we need a simulation solution for all designs
% 3. Mlir, is a compiler-based and talk about why we don't make a new langauge. why not a specific lanaguage
%     mlir has complete infrastracture, has multi-level abstractions, create a bridge bwteen the layers.
%     layers -> simulation at different level of abstraction.
% 4. overview. language -> compiler
% IN the end mention how most of simulators we mentioned dom't provide users contro; over the intermediate levels.
% continue with MLIR

% Expressing hardware has been a long time focus.

% Custom, domain-specific
Hardware accelerators are a central tool for improving efficiency in the post-Moore era.
Successful accelerators cannot be designed in a vacuum:
realizing their full potential requires simultaneous advances in algorithms and compilers.
% to support high-level programming.
% Critically,
Co-design between hardware and its accompanying software stack requires a way to rapidly simulate a proposed hardware accelerator before finalizing its design.

However, standard approaches to hardware simulation, can impede this kind of \emph{rapid iteration}.
For instance, although RTL simulation~\cite{kim2016strober, kim2017evaluation,lockhart2014pymtl,schuiki2020llhd} is valuable when the hardware is being finalized at the \emph{end} of the design process, it tends to be too detailed and too slow to be practical for earlier hardware--software co-design phases.
Designers often build custom \emph{high-level} simulators~\cite{samajdar2018scale, datta2009optimization, choi2010model, choi2019cnn}
for specific applications and architectures that sacrifice accuracy for greater flexibility and faster simulation times.
% I think we've made the point and there's no need to call out one example. --A
% For instance, SCALE-Sim~\cite{samajdar2018scale} build a systolic array simulator for convolutions.
However, these custom simulators specialize for a specific architecture model---changing the modeled hardware, such as introducing a new level in a memory hierarchy, requires rewriting substantial parts of the simulator.
\emph{General-purpose} simulation frameworks, in contrast, tend to focus on processor-centric architectures like CPUs and GPUs~\cite{power2014gem5gpu,shao2016co,binkert2011gem5},
meaning that modeling a custom accelerator architecture still requires a custom specialized implementation of the simulation logic.
Finally, traditional simulators do not expose an intermediate representation, so it can be challenging to integrate them with a compiler stack to measure performance on real software.

% %
% As a result, designers often build custom higher-level simulators~\cite{samajdar2018scale, datta2009optimization, choi2010model, choi2019cnn} that sacrifice accuracy for greater flexibility and faster simulation times, helping to facilitate algorithmic and compiler exploration.
% %
% Although they may lack perfect cycle accuracy, these simulators can still help guide coarse-grained hardware design by answering questions about bottlenecks and utilization.
% %
% In order to achieve high performance, custom C or C++ simulator implementations are often specialized for particular problems.  Adding parameters or significantly changing the architecture being modeled often takes a significant amount of low-level coding.  This is a significant barrier when the architecture under consideration is changing rapidly.  For instance, when applying AutoML techniques to automatically explore architectures for a machine learning system.
% can require rewriting substantial parts of the simulator.
%In addition, these simulators rarely integrate with compiler toolchains, requiring manual tweaking to implement a given application.
% \xxx[as]{Something about standard simulators being black boxes that are not actually programmable with compilers... integrating with a compiler stack is critical for evaluating what an accelerator can actually do for real software.}

This paper presents a general framework for rapidly implementing high-level simulators for arbitrary hardware accelerators. This framework shares basic discrete-event semantics with many existing simulation systems\cite{ptolemaeus2014system,varga2001discrete} and focuses on implementation in a multi-level compiler infrastructure, MLIR~\cite{lattner2020mlir}. This implementation enables rapid iteration and efficient, low-effort simulation of generated architectures and is intended to exist as part of an end-to-end toolchain, rather than as a standalone simulation framework.

% We aim to provide a better alternative to constructing one-off cycle-level simulators that enable rapid hardware--software co-design before the hardware design is finalized.
%The key idea is to separate a generic \emph{simulation} engine from a flexible \emph{representation} of the simulated hardware.
%
% One-off simulators conflate the hardware architecture itself with the mechanism for simulating it, making it difficult to change the simulated architecture or to configure it for a given application.
%
%We see an opportunity to leverage modern compiler technology to design a flexible representation for accelerators that enables rapid iteration and efficient, low-effort simulation.

%Our approach builds on an existing compiler infrastructure, MLIR~\cite{lattner2020mlir}, to construct an intermediate representation for simulated architectures.
The system has two main components: an MLIR \emph{dialect} for representing hardware accelerators, and a generic simulation engine that interprets those representations.
Our core contribution is an \emph{event queue} (EQueue) dialect in MLIR:
an intermediate language that represents accelerators at many levels of detail, from simple first-order models to detailed, multi-component simulations.
The dialect focuses on expressing memory allocation, data movement, and parallel execution with event-based control.
%
%After writing an EQueue specification,
%designers get simulation ``for free'':
%
The dialect is directly executable by a generic timed discrete-event simulation engine, enabling arbitrary architectures to be executed. The simulation can provide estimates of overall execution time taking into account data dependencies and resource limitations.
The simulation engine also produces visualizable operation-level traces for detailed performance analysis.

% pragraph1:

% Introduce how accelrator becomes popular
% GPU -> dedicated hardware accelerator

% paragraph2:

% 1. use of manufacture specific software pros: fast, accurate, cons: no protablity at all\\
% 2. the emergence of generalized accelartors to solve this problem, but fall into several categrories\\
% 2.1 scam sim targeting a small set of backend pro: fast, accurate, cons: not extensivle to all designs\\
% 2.2 targeting all designs, you either sacrifice performance protabiliy, or have to go down to the very low level. \\
% 2.3 domain-specific language. learning curve too high\\
% tweak many parameters/ learn the workflow of the tool
% workflow includes to high level to RTL, share same AST, type system, no need to worry about switching vetween IR very extensible

% 1. very domain-specific  new language-> need to learn new flow (the last one)
% 2. scale-sim a simulator, but only very specific workflow
% 3. RTL needs to specify level simulation. low-level on how wires are connected. early exploration high expenses, small change -> 
\begin{figure}
  \centering
  % include second image
  \includegraphics[width=0.75\linewidth]{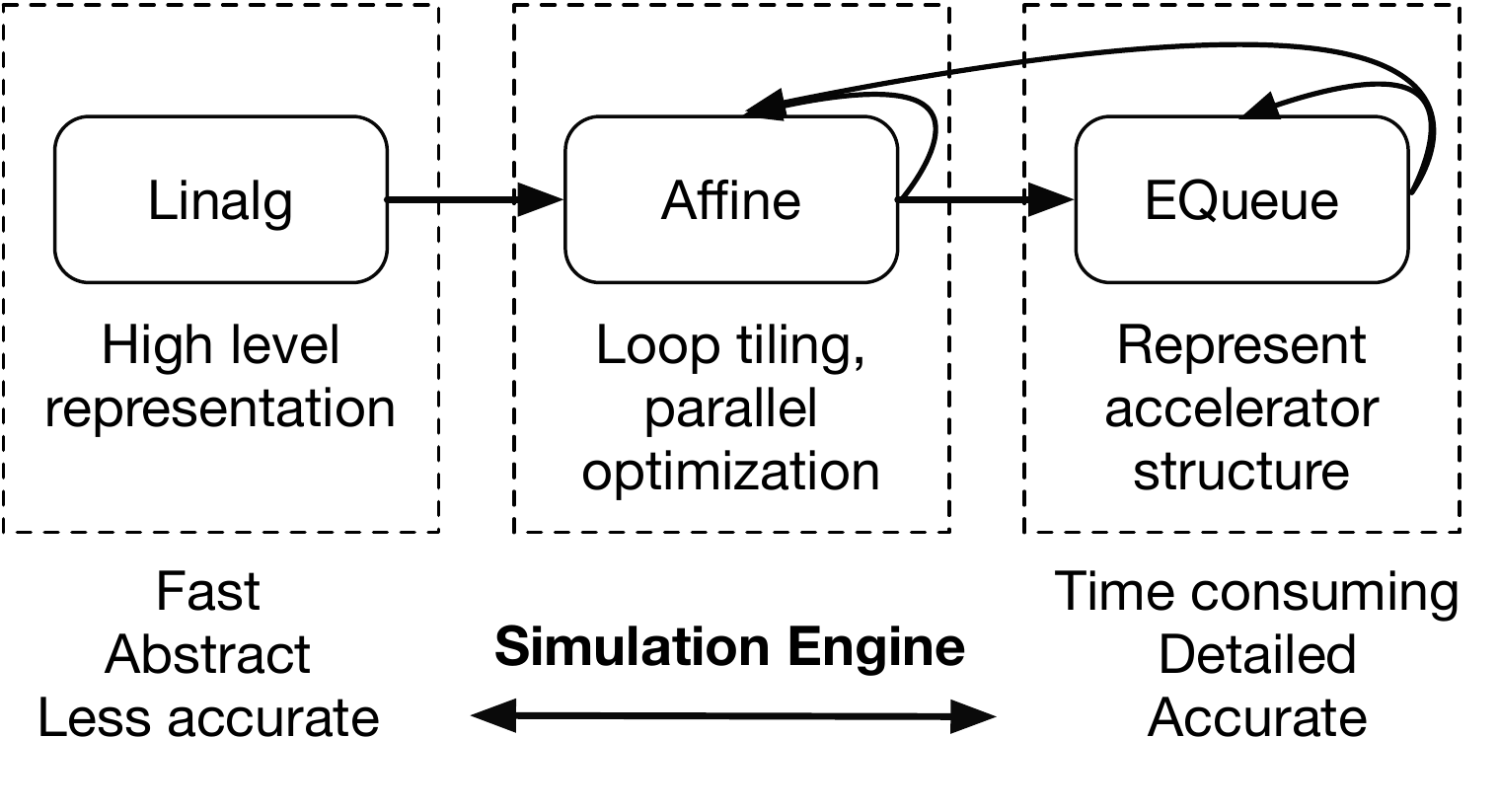}  
  \caption{An example of EQueue simulation on different levels of hardware abstraction.}
 \label{fig:mlir-overview}
\end{figure}

By building on MLIR, the \equeue\ leverages a broad ecosystem of transformations, analyses, and other dialects.
%
% By building on a software compiler infrastructure, we enable analysis and transformation tools that are not possible with one-off implementations.
% Developers can easily write passes that compile higher-level programs into accelerators
% or transform the architecture.
% By building on MLIR, the EQueue simulation dialect can directly leverage a broad ecosystem of other dialects and tools.
Designers can quickly prototype compilers from high-level languages to
% Referencing "high-level languages" will satisfy the reviewers who asked specifically about how you are supposed to program these things. --A
% abstractions where the accelerator is viewed as black box to
lower-level accelerator configurations.
As a software compiler infrastructure, MLIR also enables analysis and transformation tools that are not possible with one-off simulators.
For example, Fig.~\ref{fig:mlir-overview} shows a lowering pipeline for progressive optimization of tensor computations using existing MILR dialects:
the high-level \emph{Linalg} dialect to represent tensor operations,
the \emph{Affine} dialect to express explicit loop tiling,
and finally our new EQueue dialect to model explicit data movement among hardware components.
Critically, such a lowering pipeline enables simulation at multiple levels of detail:
users can get quick-and-dirty performance estimates at the Linalg level on tensor behavior,
or they can lower gradually to more detailed EQueue hardware simulations for more accurate but costly estimation.

Compared to traditional one-off accelerator simulators,
we see several advantages in the compiler-based approach:
\begin{enumerate}
    \item We can simulate at different points in a compilation flow, representing a hardware at multiple abstraction levels.
    \item Changes to the architecture are decoupled from changes to the simulation logic, so design iteration can be easier.
    \item By reusing compiler passes with different parameters to transform the architecture, designers can easily switch among different architectures for the same computation.
\end{enumerate}
This paper presents the \equeue\ and its open-source\footnote{https://github.com/cucapra/EventQueue} implementation using MLIR.
We explain the core constructs via a running example (\cref{sec:overview}) and then detail the dialect (\cref{sec:language}) and its simulation engine (\cref{sec:simulation}).
We show the flexibility of programming with two case studies: a systolic array for deep learning computations ,
and a model of the AI Engine cores in Xilinx's Versal ACAP fabric.
%In both cases, our goal is not to beat existing simulator in performance, but to illustrate it capability to express various architecture with low cost.
In the first case,
we compare our EQueue-based simulator against a traditional, special-purpose systolic array simulator,
SCALE-Sim~\cite{samajdar2018scale}.
The EQueue approach matches its accuracy while offering better flexibility to rapidly change the modeled data flow~(\cref{sec:systolic}).
In the second case, we demonstrate how EQueue's flexibility can guide designers to improve their designs on a real-world reconfigurable architecture~(\cref{sec:aie}).
% We also describe two case studies of accelerator simulations we built using the \equeue:
% a standard systolic array for deep learning computations (\cref{sec:systolic}),
% and a model of the AI Engine cores in Xilinx's Versal ACAP fabric~(\cref{sec:aie}).
% In the first case, we demonstrate that EQueue-based simulation matches the results to a state-of-the-art simulator while lowering the overhead to switch between dataflows.
% In both cases, we show that the \equeue\ enables flexible explorations of hardware--software co-design questions.

\section{Overview by Example}
\label{sec:overview}

\begin{figure*}[t]
\begin{minipage}{0.78\textwidth}
\begin{subfigure}{\textwidth}
  \centering
  % include first image
  \includegraphics[width=\linewidth]{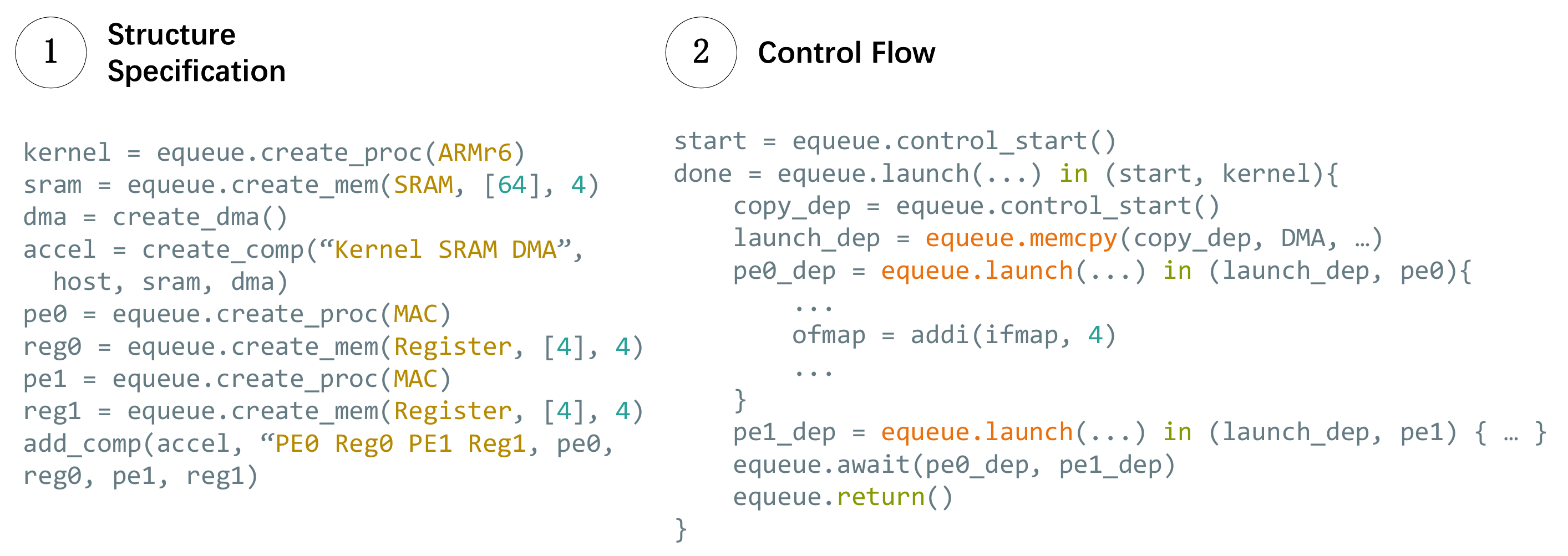}
  \caption{An EQueue program as input to our generic simulator.}
 \label{fig:overview-example-program}
\end{subfigure}
\end{minipage}\hfill
\begin{minipage}{0.2\textwidth}
\begin{subfigure}{\textwidth}
  \centering
  % include second image
  \includegraphics[width=\linewidth]{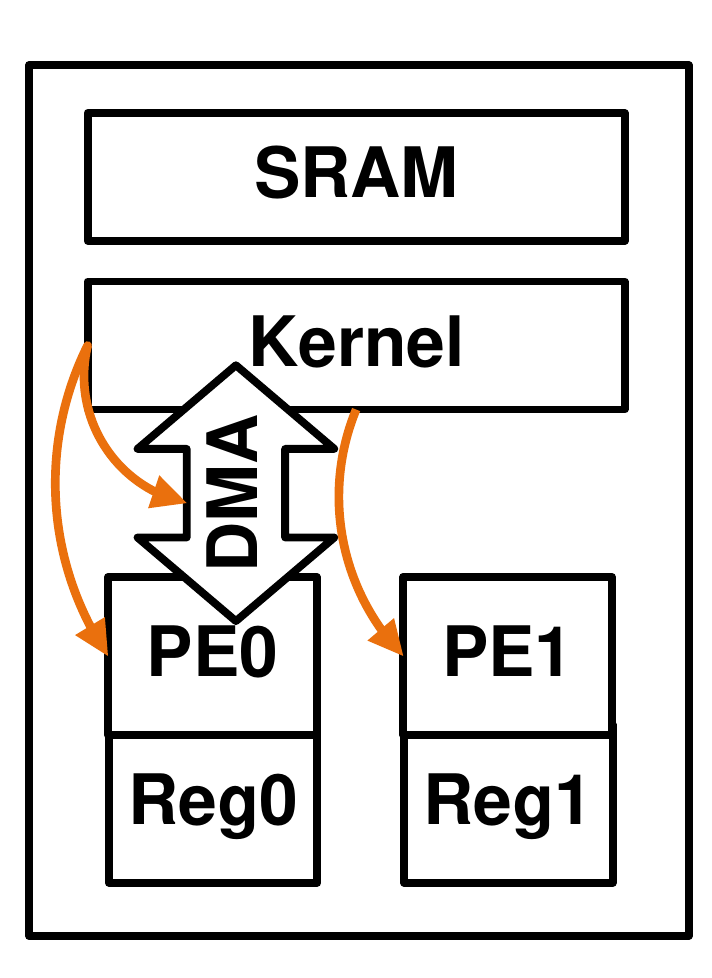}  
  \caption{Accelerator.}
 \label{fig:overview-example-host}
\end{subfigure}
\end{minipage}
\caption{Modeling an accelerator with an EQueue program and the model created by the simulation engine. Code listings omit types and the \% prefix for legibility.}
%The event queue of host processor omitted.
 \label{fig:overview-example}
\end{figure*}

% \begin{figure*}[t]
% \begin{minipage}{0.49\textwidth}
% \begin{subfigure}{\textwidth}
%   \centering
%   % include first image
%   \includegraphics[width=\linewidth]{fig/overview-example-program.pdf}
%   \caption{Simplified EQueue program for an example accelerator.}
%  \label{fig:overview-example-program}
% \end{subfigure}
% \end{minipage}\hfill
% \begin{minipage}{0.49\textwidth}
% \begin{subfigure}{\textwidth}
%   \centering
%   % include second image
%   \includegraphics[width=0.9\linewidth]{fig/overview-example-host.pdf}  
%   \caption{Initial control flow of the example accelerator.}
%  \label{fig:overview-example-host}
% \end{subfigure}\hfill
% \begin{subfigure}{\textwidth}
%   \centering
%   % include second image
%   \includegraphics[width=0.9\linewidth]{fig/overview-example-pes.pdf}  
%   \caption{Optimized control flow of the example accelerator.}
%  \label{fig:overview-example-pes}
% \end{subfigure}
% \end{minipage}
% \caption{Mapping a summation operation on an architecture with \equeue\ and its perspective from the generic simulation engine. Code listings omit types and the \% prefix for legibility.}
% %The event queue of host processor omitted.
%  \label{fig:overview-example}
% \end{figure*}

This section summarizes our simulation flow. We write an EQueue program (Fig.~\ref{fig:overview-example-program}) and show how our simulation engine executes it.
Fig.~\ref{fig:overview-example-host} depicts the modeled architecture.
% \cref{sec:language,sec:simulation} then describe them in complete detail.

\subsection{Structure Specification}
%An EQueue-structured MLIR program is composed of \circled{1} a structure specification and \circled{2}\circled{3} a control flow. 

% The \equeue\ describes hardware accelerators, including the way they share structural resources and move data.
%The key idea to EQueue-structured program is modeling explicit data movement in the context of shared resources and operation dependencies arbitrary architecture according to hardware designer's description. 
% An architecture is specified with a structure description and control flow. 

The first main part of an EQueue program is a set of \emph{structural} declarations, which define the hardware resources that make up an accelerator.
% An EQueue program specifies an architecture by describing
% the way it shares structural resources and moves data.
Fig.~\ref{fig:overview-example-program} lists an EQueue program describing a toy accelerator in Fig.~\ref{fig:overview-example-host}.
% The code is semi-pseudo in the sense that we ignore variable prefix \% and types for the ease of readability.
First,
% EQueue programs specify the architecture with
\code{create_*} operations instantiate components like processing elements (PEs) and memories.
Then, \code{launch} operations map work onto this structure to specify the computation.
% they map operations onto this structure using \code{launch} operations to specify the computation.

We start with structure specification~\circled{1}. The program uses several \code{create_*} operations to declare components including processors, memories, and direct memory access (DMA) units.
These \code{create_*} operations select from a range of primitive component types (\code{ARMr6}, \code{SRAM}, \code{Register}, etc.).
These tags correspond to performance models in the simulation engine:
for example, the simulation model for \code{SRAM} components has slower warm-up time, slower reads, and higher power usage than the \code{Register} model.
Programs can assemble
these components into hierarchies using
\code{create_comp} to create a new component and \code{add_comp} to add one to an existing component.

\subsection{Control Flow}

% Every operation in a launch region is \emph{sequentially} executed, but we also want to model multiple processors executing concurrently. The solution is to model \code{launch} as \emph{event operation}, which is pushed to the \emph{event queue} when simulator encounters it.
The second part of a EQueue program is its control flow.
The core operation is \code{launch}.
The \code{launch} operation takes a dependency, a processor component, and a block of code.
The simulation engine implements \code{launch} by issuing code blocks on processors, which execute sequentially.
\xxx[as]{I think this is too vague---people won't know what an "operation iterator" or a "time log" is. Just delete this next sentence?}
%It tracks the execution status with an operation iterator, time logs, etc.

%There are exceptions. 
\code{launch} or \code{memcpy} are event operations executed out of order.
\xxx[as]{I think it's confusing to say that these are exceptions. They still execute sequentially, just like all other operations in a code block. It's just that their effect is to enqueue some \emph{other} block to run later. Let's not say they are exceptions and instead just directly say what they do.}
Every processor in the simulation has an \emph{event queue}; launching a code block enqueues it for the given processor.
% When encountering these operations, the simulation engine enqueues them to event queues. They are structures the engine creates for every processor.
At simulation, the engine checks the dependency and executes code blocks when their dependencies are ready.
Processors communicate by spawning events with \code{launch} or \code{memcpy}.

For example, the control flow~\circled{2} illustrates how the \code{Kernel} processor distributes its work to \code{DMA}, \code{PE0}, and \code{PE1}. Fig.~\ref{fig:overview-example-host} indicates the communication using arrows.
\code{DMA}, \code{PE0}, and \code{PE1} are all independent processors but can communicate through their event queues.
As the \code{launch} of \code{PE0} and \code{PE1} both depend on \code{mempcy} operation of \code{DMA}, \code{PE0} and \code{PE1} start simultaneously.

% We can modify our summation example using parallelism.
% An updated control flow description~\circled{3} splits the summation across two PEs, and Fig.~\ref{fig:overview-example-host} illustrates the accelerator's data movement.
% The \code{host} processor sends half of the addition to \code{PE0} and the other half to \code{PE1} with \code{launch} operations.
% The \code{launch} operations are pushed to the event queues of \code{PE0} and \code{PE1}, so the two processors run in parallel.
%Therefore though \code{launch} and \code{memcpy} on \code{PE0} can take time to execute, they do not block the execution of \code{PE1}'s \code{memcpy} and \code{launch}.% This way they do not block other operations that might be ready to issue.
% The simulation engine constantly checks the head of the event queue of each processor to see if its dependent event finishes execution. In this example, \code{memcpy} depends on the event \code{start_copy}.
%an event operation is ready to issue.

% Hoping the below paragraph captures the same idea as the commented-out one below while saving space. --AS
\noindent\textbf{Benefits.}
This example shows how EQueue describes accelerator structure and control flow \emph{separately} from the simulation logic.
Designers can change the architecture without needing to modify the simulation engine, which reduces the cost of exploring alternative designs.
It also shows how EQueue programs can intermix code from other MLIR dialects for core accelerator logic:
namely, the \code{addi} operation is from MLIR's standard dialect.
The MLIR ecosystem offers operations from many abstraction levels, from linear algebra to machine code.
These help to express architectures at different levels, from abstract black boxes to low-level implementation details.

\section{EQueue Dialect}
\label{sec:language}

We illustrate the \equeue\ with four parts: modeling hardware, expressing data movement, launching computation and concurrency between controllers.
% We illustrate the three features with example programs.
%For ease of readability, we omit the \% prefix and type annotations.

\begin{figure*}[ht]
\begin{minipage}{0.32\textwidth}
  \centering
  % include first image
  \includegraphics[width=0.5\linewidth]{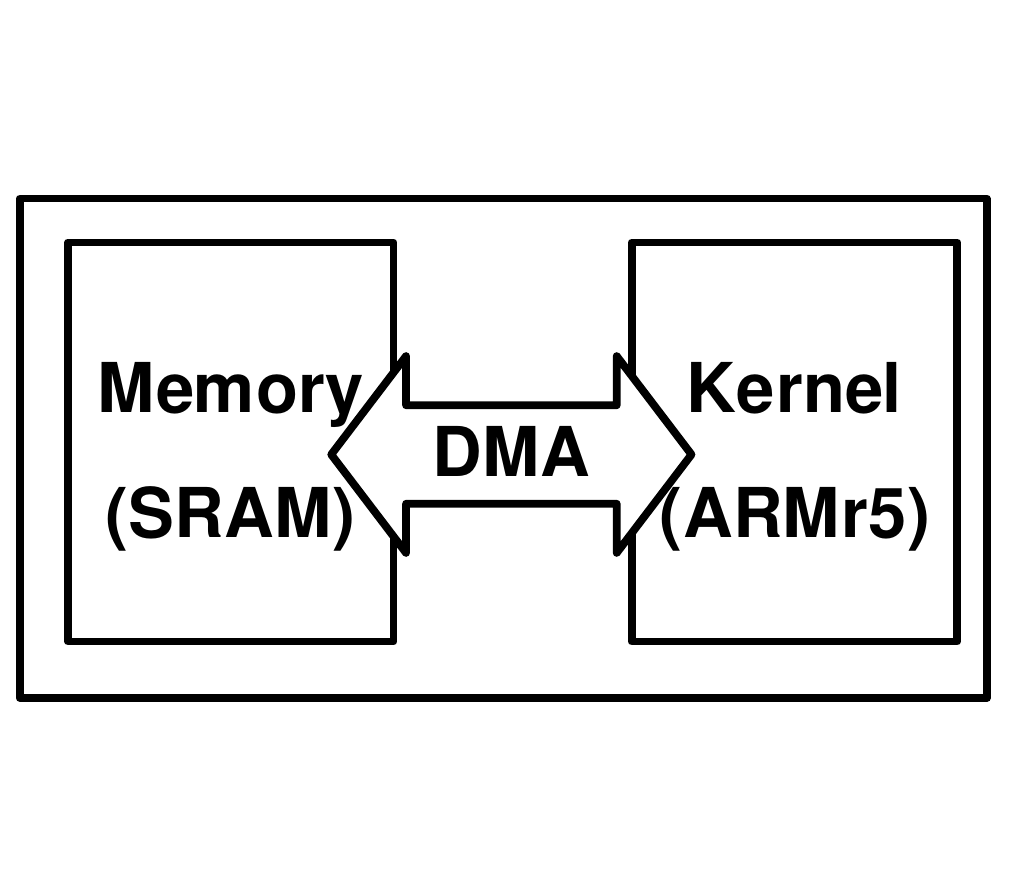}  
\caption{Simple one-core accelerator.}
\label{fig:one-core}
\end{minipage}\hfill
\begin{minipage}{0.66\textwidth}
\begin{subfigure}[t]{0.49\textwidth}
  \centering
  % include first image
  \includegraphics[width=0.5\linewidth]{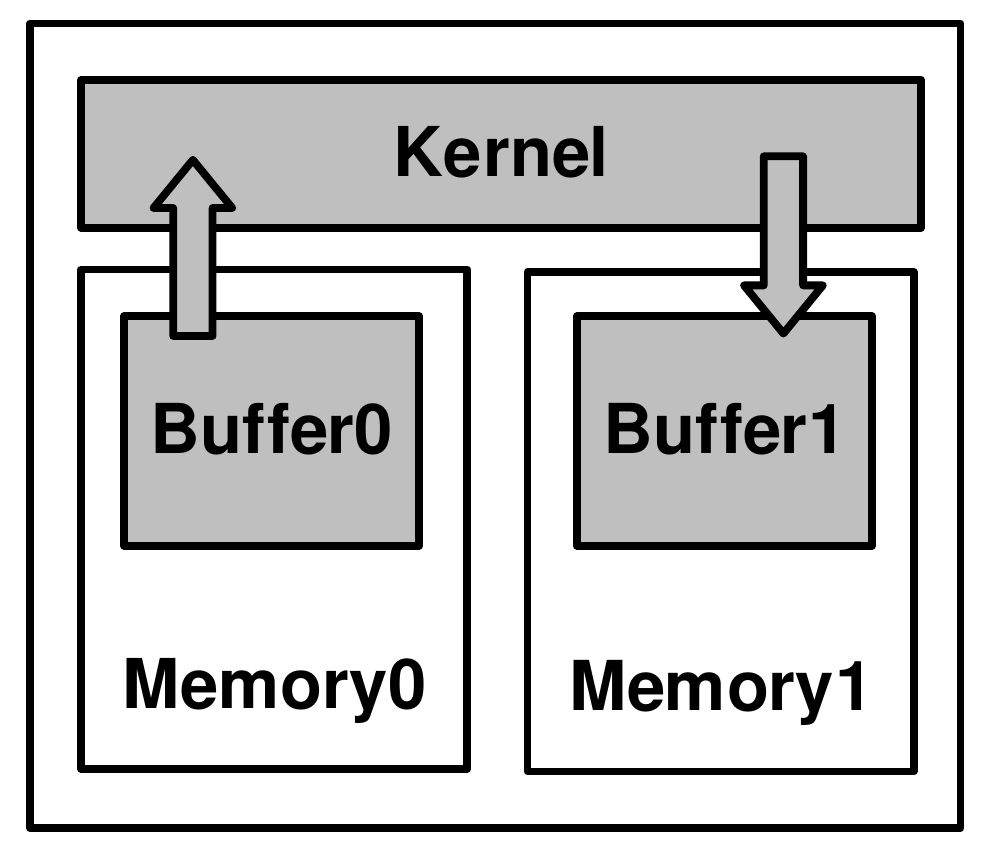}  
  \caption{Data movement controlled by kernel.}
 \label{fig:kernel-memcpy}
\end{subfigure}
\begin{subfigure}[t]{.49\textwidth}
  \centering
  % include second image
  \includegraphics[width=0.5\linewidth]{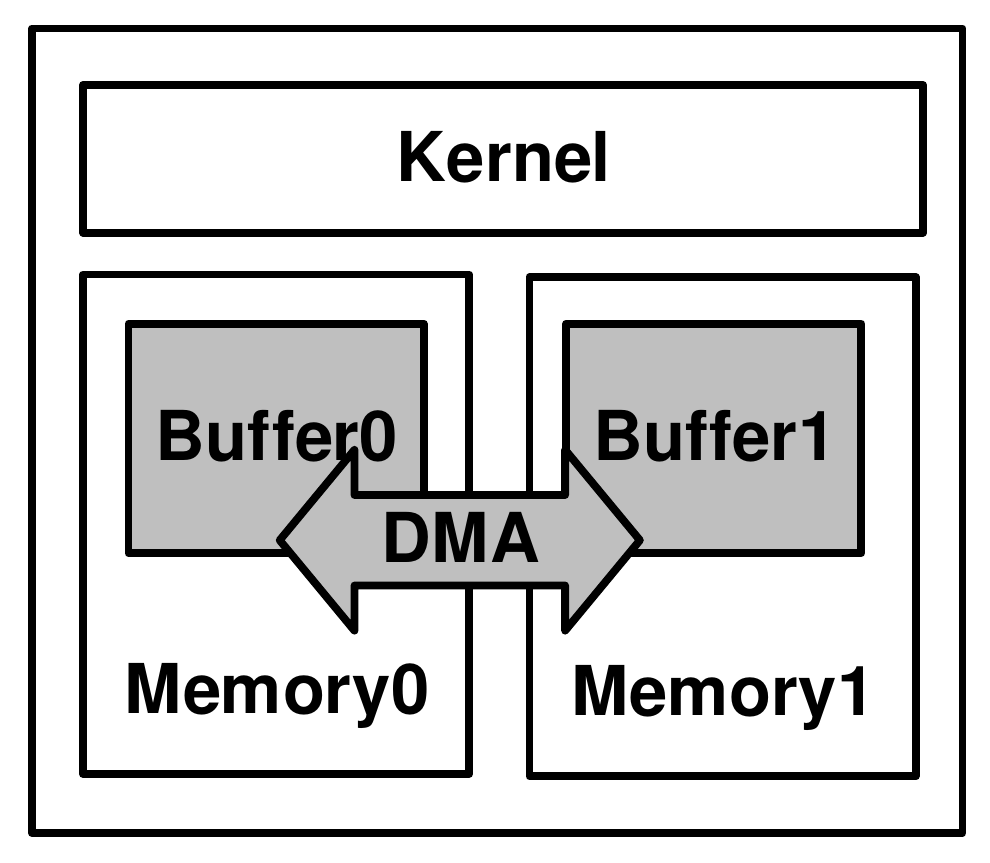}  
  \caption{Data movement controlled by DMA.}
 \label{fig:dma-memcpy}
\end{subfigure}
\caption{Expressing explicit data movement using \equeue.}
\label{fig:memcpy}
\end{minipage}\vfill
\begin{minipage}{\textwidth}
\begin{subfigure}[t]{0.3\textwidth}
  \centering
  % include first image
  \includegraphics[width=\linewidth]{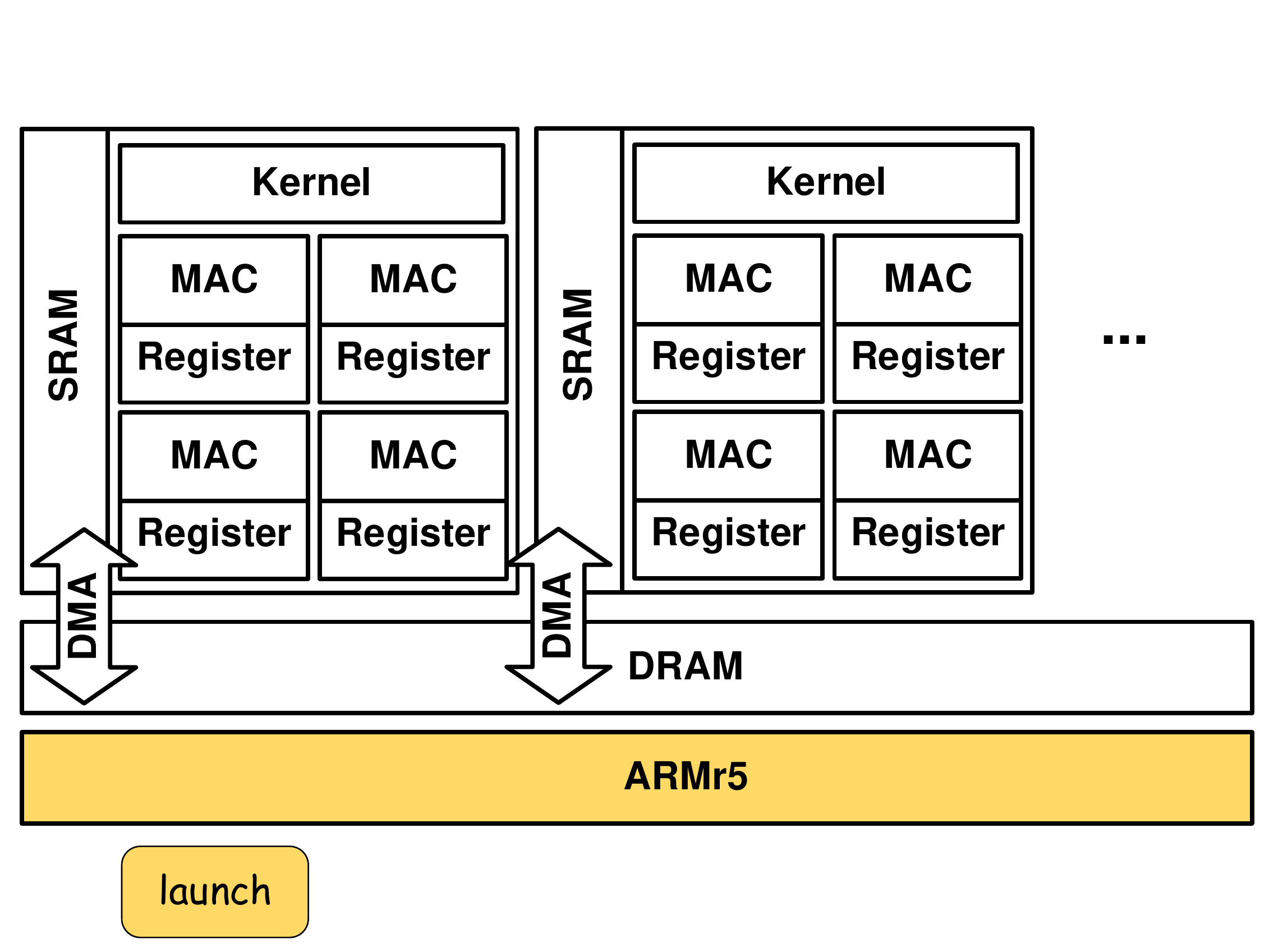}  
  \caption{Event queue of ARMr5 processor when \code{launch} waits on its dependency.}
 \label{fig:concurrency-step1}
\end{subfigure}
\begin{subfigure}[t]{.35\textwidth}
  \centering
  % include second image
  \includegraphics[width=\linewidth]{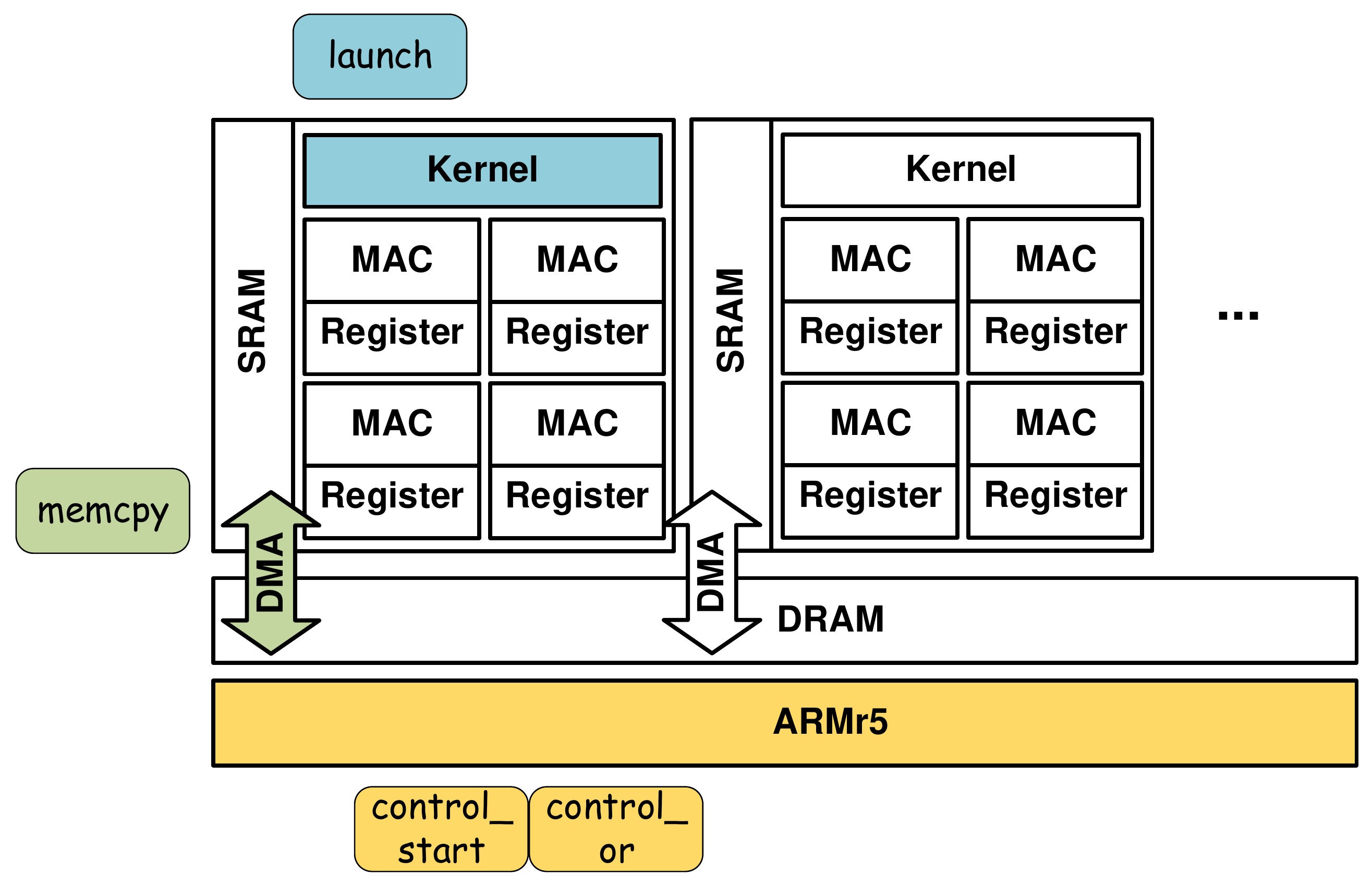}  
  \caption{Event queues of Kernel and DMA when \code{launch} and \code{memcpy} wait on their dependency.}
 \label{fig:concurrency-step2}
\end{subfigure}
\begin{subfigure}[t]{.3\textwidth}
  \centering
  % include second image
  \includegraphics[width=\linewidth]{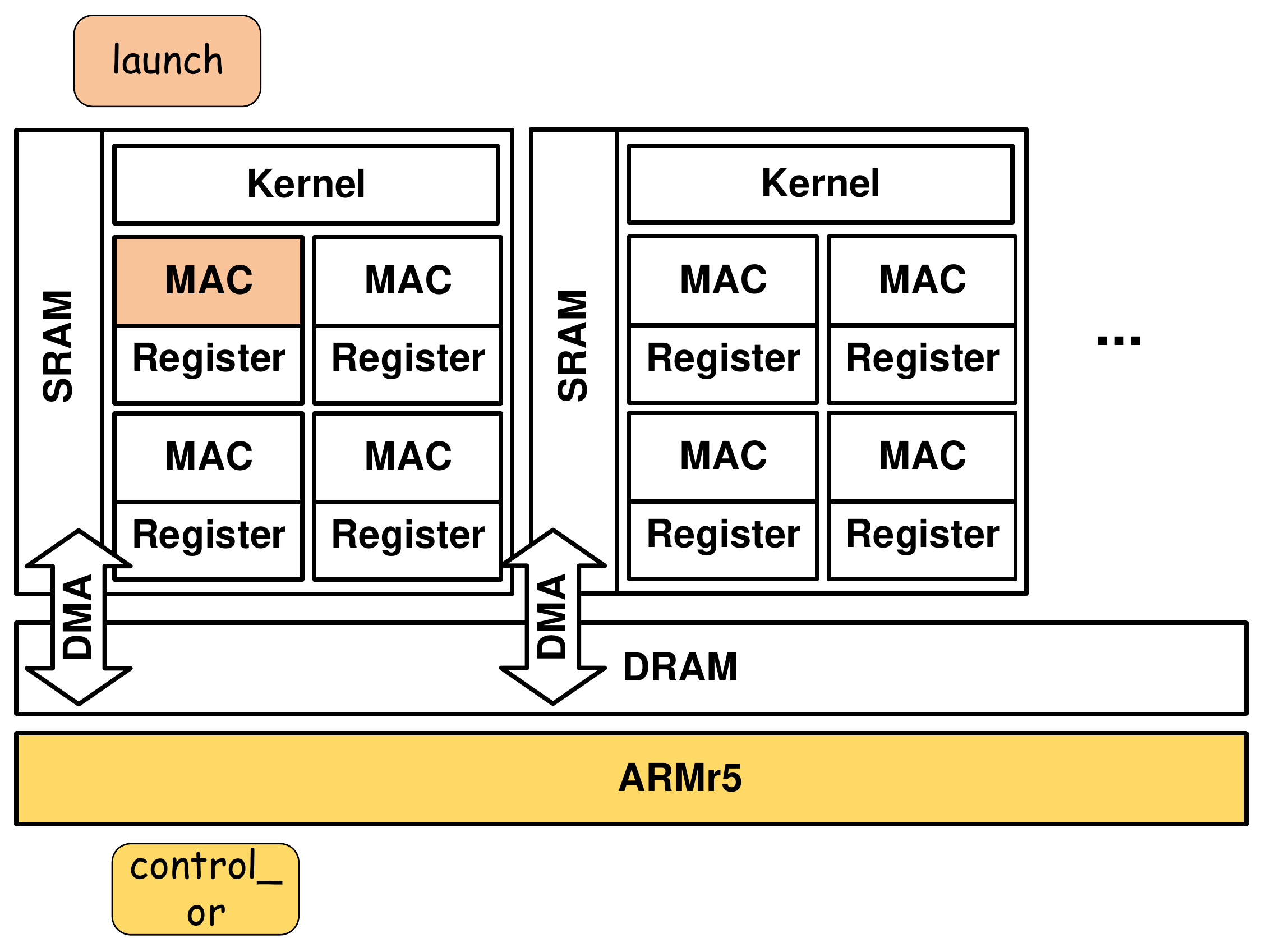}  
  \caption{Event queue of MAC when \code{launch} wait on its dependency.}
 \label{fig:concurrency-step3}
\end{subfigure}
\caption{Three stages of the timeline of execution of the accelerator in \cref{fig:concurrent_code}.}
 \label{fig:concurrency}
\end{minipage}
\end{figure*}

\subsection{Modeling Structure}
\label{sec:language-structure}

The \equeue\ lets programs declare structural components to model hardware.
For example, this code creates the structure from Fig.~\ref{fig:one-core}:
\begin{lstlisting}
kernel = equeue.create_proc(ARMr5)
mem = equeue.create_mem([4096], 32, 4, SRAM)
dma = equeue.create_dma()
accel = equeue.create_comp("Memory Kernel DMA", 
mem, kernel, dma)
\end{lstlisting}
There are three kinds of components:
processors, memories, and DMA engines.
A processor is a component that can execute commands, via the \code{launch} operation described in~\cref{sec:language-concurrency}. A DMA component is a specialized processor that is only used for data movement.
A memory stores data; its speed is affected by its type, size, and ports.
Each \code{create_*} operation encodes component properties in its arguments.
For instance, \code{create_mem([4096], 32, 4, SRAM)} declares a memory component of SRAM type with 4 banks and 4096 data elements of 32 bits each.

The \code{create_comp} composes smaller components into a hierarchy of larger components.
The example code declares a component \code{accel} with three subcomponents with the names ``Memory", ``Kernel" and ``DMA".
%\comment{what do we gain by grouping processor, memory, dma component, what cannot be tried if they are separate}
Later, code can look up components in the hierarchy using a \code{get_comp} operation:
\begin{lstlisting}
dma = equeue.get_comp(accel, "DMA")
\end{lstlisting}

% One can also add advanced components to an existing component with \code{add_comp} and \code{get_comp}. 
% \begin{lstlisting}
% connection = equeue.create_connection(Streaming, 32)
% equeue.add_comp(accel, "connection", connection)
% connection_get = equeue.get_comp(accel,
%     "connection")
% \end{lstlisting}

Finally, \emph{connections} model bandwidth constraints:
\begin{lstlisting}
connection = equeue.create_connection(Streaming, 32)
\end{lstlisting}
The \code{create_connection} operation has two arguments: their type and their bandwidth in bytes per cycle.
The two types are \code{Streaming}, which allows simultaneous reads and writes, and \code{Window}, which models a buffer that requires locking for exclusive access.
Streaming interfaces typically offer lower latency while windowed interfaces offer higher bandwidth.
%We model connections as either streaming that allows constant reading and writing or windows interface which requires a lock while providing large bandwidth. 
The simulation engine outputs profiling statistics for each connection's bandwidth utilization over time.
The bandwidth limit is optional; the simulation engine can also model infinite-bandwidth connections and still collect statistics, as we show in~\cref{sec:aie-case1}.
%Bandwidth is also an optional argument. If it is not declared, the simulation engine assumes unlimited accessing speed. It can be used to profile actual bandwidth, as we will show in Section 6-E.
%Default reading and writing operations in \equeue\ assume no bandwidth constraint, which can be helpful at the starting point to design a system, as we will show in Section 6-E.

%Since components can be grouped through \code{create_comp} and \code{add_comp}, one can model arbitrary hardware hierarchy. %Fig.~\ref{fig:large-scale-accel} shows a large accelerator that can be created with the hardware modeling commands.

\subsection{Explicit Data Movement}
\label{sec:language-data-movement}

Given the hardware structure, we can specify data movement using allocation, deallocation, and read/write operations on memories.
For example, consider two memories:
\begin{lstlisting}
mem0 = equeue.create_mem([4096], 32, 4, SRAM)
mem1 = equeue.create_mem([4096], 32, 4, SRAM)
conn = equeue.create_connection(Streaming, 32)
\end{lstlisting}
We will model the data movement
shown in Fig.~\ref{fig:memcpy}.
To associate a buffer with a memory, we use \code{alloc}:
\begin{lstlisting}
buffer0 = equeue.alloc(mem0, [64], 32)
buffer1 = equeue.alloc(mem1, [64], 32)
\end{lstlisting}
The \code{alloc} operation specifies the memory, buffer size in elements, and element size in bits.
% Fig.~\ref{fig:kernel-memcpy} illustrates these buffers.

To model data movement, we use \code{read} and \code{write} operations to transfer data into a buffer, optionally through a connection:
\begin{lstlisting}
data = equeue.read(buffer0, conn)
equeue.write(data, buffer1, conn)
\end{lstlisting}
Both operations take a buffer and, optionally, a connection; \code{write} also takes the value to write.
Here,
we use connection whose bandwidth is 32 bytes per cycle.
Finally, programs use \code{dealloc} to free buffers:
\begin{lstlisting}
equeue.dealloc(buffer0)
equeue.dealloc(buffer1)
\end{lstlisting}
So far, these operations specify how data moves, but not the processors executing the operations.
The next subsection shows how to assign this code to processors.

\subsection{Launching Computations}
\label{sec:launching_comp}

The \code{launch} operation schedules blocks of code onto processors.
This code runs a block on the \code{kernel} processor:
\begin{lstlisting}
equeue.launch(buffer0, buffer1 = b0, b1)
  in (kernel){
    data = equeue.read(buffer0)
    equeue.write(data, buffer1)
    equeue.return
}
\end{lstlisting}
The arguments to \code{launch} pass resources that the code block, represented by an MLIR region, may access.
The code in the region will be dispatched to the particular processor for execution. When the region is executed, the resources will be available, enabling the region to run to atomically run to completion. Although most The code runs sequentially.
Fig.~\ref{fig:kernel-memcpy} illustrates the above data movement.
%We will discuss more how \code{launch} operation works in \cref{sec:language-concurrency}. 

The \code{memcpy} operation is syntactic sugar for a \code{launch} that reads and then writes data.
\code{memcpy} is mostly used with DMA units.
% as its main contribution is to manage data movements.
Fig.~\ref{fig:dma-memcpy} shows the data movement in this code:
\begin{lstlisting}
equeue.memcpy (mem0, mem1, dma)
\end{lstlisting}
Launching a code block enqueues it for later execution; the next section describes how this queueing work.

\subsection{Concurrent Event Scheduling}
\label{sec:language-concurrency}

Concurrency in the \equeue\ occurs at the granularity of \emph{events}.
While the code within a \code{launch} block executes sequentially, it can use \emph{event operations} to spawn asynchronous, concurrent work.
% As \cref{sec:language-structure} describes, a processor is a component that can launch and execute code blocks.
% % , while running in parallel with other components. A DMA component is a special processor that only issues \code{read} and \code{write} operations. \code{memcpy} is the syntactic sugar for one \code{read} followed by one \code{write}.
% The code inside a \code{launch} blocks execute sequentially, but it can also create \emph{event operations} that execute asynchronously.
Event operations include \code{launch} and \code{memcpy} described above,
and also logical operations on events: \code{control_start}, \code{control_and} and \code{control_or}.

Event operations can have \emph{dependencies}, indicating other events they depend on that must finish before the event can start.
%Every event operation generates one \emph{dependency}.
\code{launch} and \code{memcpy} each have one dependency, and
\code{control_start} has none: it is a special operation for beginning a chain of events.
\code{control_or} and \code{control_and} are ready when any or all of their dependencies finish, respectively.

During simulation, launching an event pushes it onto a given processor's \emph{event queue}.
Different processors can execute events from their queues in parallel, but each processor only executes one event at a time.
Events can launch other events, so simulations can nest \code{launch} operations in arbitrary ways to reflect their control hierarchy.

The \equeue\ includes an
\code{await} operation that blocks execution until a different event completes.
Finally, a \code{launch} block can pass values out with the \code{return} operation.
% The $i$-th input argument of a \code{return} corresponds to the $i+1$-th result of \code{launch} ends with that \code{return}.
% In the program above, \code{ret} corresponds to \code{done_compute}. Notice that \code{ret} is not available till the outer-most \code{launch} operation finishes, i.e., \code{done} is generated.

\begin{figure}
\begin{lstlisting}
start = equeue.control_start()
// Event in Fig. 5(a).
done, ret = equeue.launch (buf0, buf1, dma, kernel, mac = b0, b1, d, k, m)  in (start, ARMr5){
    // Event in Fig. 5(b).
    start_event = equeue.control_start()
    done_dma = equeue.memcpy(start_event, 
        buf0, buf1, dma)
    done_kernel = equeue.launch(MAC=mac) in 
        (start_event, kernel){
        // Event in Fig. 5(c).
        start_mac = ...
        done_mac = equeue.launch(...) in 
            (start_mac, MAC){ ... }
        ...
    }
    done_compute = equeue.control_and(
        done_kernel, done_dma)
    equeue.await(done_kernel)
    equeue.return(done_compute)
}
\end{lstlisting}
\caption{Example showing concurrent execution of an ARMr5 control processor, a DMA engine, and a MAC unit.}
\label{fig:concurrent_code}
\end{figure}

\smallskip
\noindent
\textbf{Example.}
Fig.~\ref{fig:concurrent_code} shows an example accelerator that uses concurrent tasks,
and
Fig.~\ref{fig:concurrency} illustrates the timeline of its execution.
One ARMr5 processor uses a DMA unit for data transfer and a small MAC kernel using \code{launch} operations.
% The event queue lets the DMA and MAC kernels run concurrently and
The \code{control_*} operations encode the execution order while allowing parallel execution on the DMA and MAC kernel.
%
% In this program, \code{control_start}, \code{launch}, \code{memcpy} and \code{control_or} are all event operations. %\code{start_control} and \code{control_or} and are pushed to the event queue of \code{ARMr5}, \code{launch} is pushed to event queue of \code{kernel} and \code{memcpy} is pushed to that of \code{dma}. 
% During execution, the event queues keep track of the independent processors and tells them the right time to start execution.

%\xxx[as]{On second reading, I'm a little more confused by this paragraph. Things I am confused by: (1) What exactly is \code{start}? It only appears once in the code and is not used again. (2) The text seems to refer to an event called \code{launch}, which is presumably the one created by the \code{done_launch} line, but that's a confusing name because the operation is also called \code{launch}. And there are two of them. Can we call them, like, \code{event1} and \code{event2} or something to disambiguate?}

Fig.~\ref{fig:concurrency-step1} highlights the event queue of \code{ARMr5} when it waits on \code{start} event to issue \code{launch}. 
Fig.~\ref{fig:concurrency-step2} shows when \code{start} is generated, the \code{launch} is removed from the event queue of \code{ARMr5}. 
Then, \code{control_start} and \code{control_or} are pushed to the queue of \code{ARMr5}, \code{launch} is pushed to the queue of \code{kernel}, and \code{memcpy} is pushed to the queue of \code{dma}. This way, the event operations run concurrently since they do not block the execution of other operations in \code{ARMr5}'s \code{launch} block.
Finally, since \code{memcpy} of \code{dma} and \code{launch} of \code{kernel} both depend on \code{start_event}, once it finishes, both \code{memcpy} and \code{launch} can be issued from their corresponding event queues, as indicated by Fig.~\ref{fig:concurrency-step3}. 
Fig.~\ref{fig:concurrency-step3} also shows that \code{launch} of \code{MAC}, after its dependency finishes, is pushed to event queue of \code{MAC}.
The \code{return} operation passes \code{done_compute} back to the top level as the result value, \code{ret}. Notice that \code{done}, the first return value of \code{launch}, is the dependency generated by \code{launch}.

\subsection{Introducing External Operations}
\label{sec:language-external}

Sometimes there are special cases where existing MLIR dialects cannot express a specific hardware operation. We introduce \code{op} to address this situation:
\begin{lstlisting}
res0, ... = equeue.op("mac", {arg0, arg1, ...})
\end{lstlisting}
\code{op} takes in a \emph{signature} specifying the operation name and an arbitrary number of inputs and outputs. Here the signature is \code{"mac"}, which can be modeled as multiplication and addition in the one cycle in the simulator library.
% , and the inputs and outputs are \code{arg0, arg1, ...} and \code{res0, ...} correspondingly.
The simulation engine checks the signature to jump to the operation's implementation specifying cycle counts and the simulated behavior.

\section{Simulation}
\label{sec:simulation}
This section introduces the EQueue simulation engine.
Fig.~\ref{fig:simulation-flow} shows an overview of the simulation workflow.

%The simulator takes in an EQueue-structured MLIR file, simulates the hardware execution process described by the MLIR file in a loop and generates profiling summary and tracing file.

\subsection{Inputs}
\label{sec:simulation-input}
\begin{figure}[t]
  \centering
  % include second image
  \includegraphics[width=\linewidth]{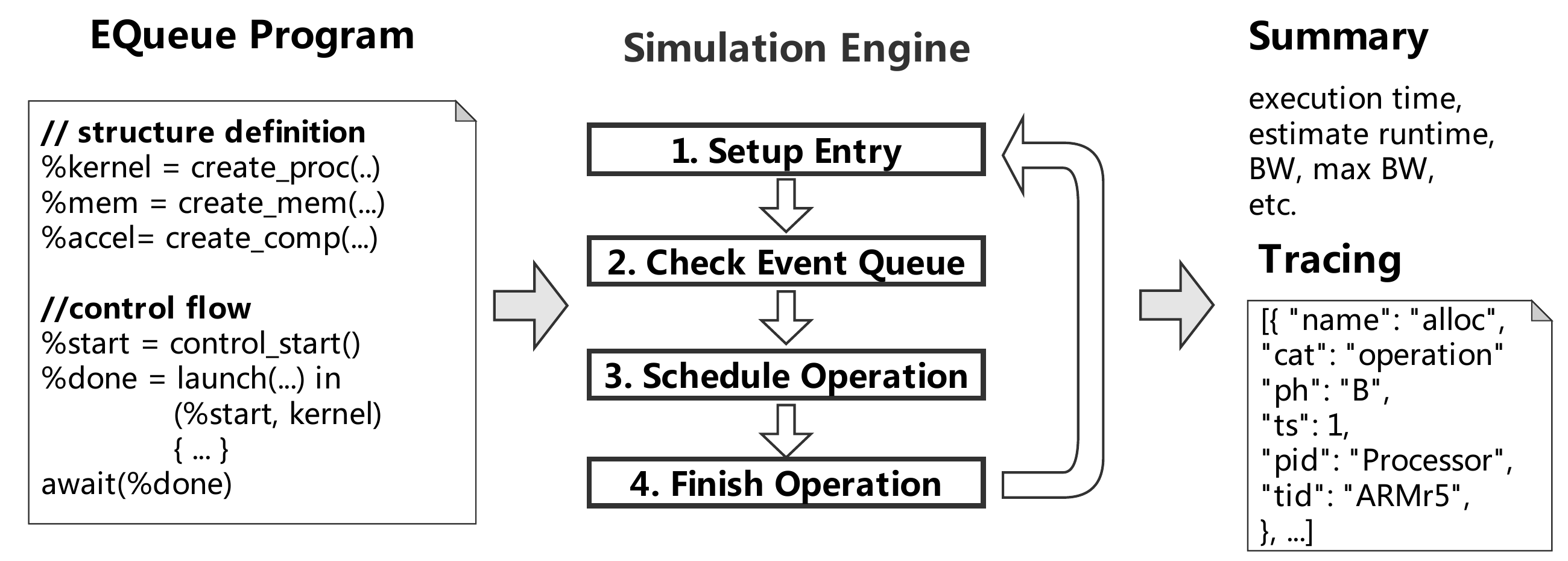}  
  \caption{The simulation workflow.}
  \label{fig:simulation-flow}
\end{figure}

%An EQueue-structured MLIR file is composed of structure specification as we introduced and control flow. The control flow is not limited to \equeue\ dialect. 
The simulation engine takes in an EQueue program.
As Fig.~\ref{fig:simulation-flow} shows, an EQueue program is composed of a structure definition and a control flow.
%The control flow can be composed of not only EQueue dialect, but Standard dialect, Linalg dialect, Affine dialect and etc.
%
Designers can produce EQueue programs by writing simple \emph{generators} in C++,
as we demonstrate in~\cref{sec:systolic-dataflow}.
Alternatively, compilers can translate to EQueue from high-level dialects such as Linalg, as we show in~\cref{sec:systolic-lowering}.
The infrastructure includes many reusable passes~(\cref{sec:passes}) to enable these lowering pipelines.
%with abstraction across multiple levels as Fig.~\ref{fig:mlir-overview} indicates.

\subsection{Outputs} 
\label{sec:simulation-output}
The simulation engine outputs a profiling summary and a visualizable tracing file.
The profiling summary includes the simulation execution time, the simulated runtime in cycles, read and write bandwidth for each connection, maximum bandwidth, and the total bytes read or written for each memory.
We also report a \emph{max bandwidth portion} for both read and write bandwidth, which is the fraction of the total simulated runtime spent at a channel's maximum bandwidth.
% The max portion tell how long time the maximum read or write lasts. 
The designer can use this statistic to adjust bandwidth accordingly to avoid waste or increase computation utilization.
%to the portion is low or reduces bandwidth to avoid waste when the bandwidth is high.

The trace is a JSON file with operation-wise records in event trace format~\cite{trace-event-format}. %It logs every operation start and end time with its corresponding processor and memory components. 
The Chrome browser can visualize this event trace format~\cite{chrome-tracing}.
We show in~\cref{sec:aie} on how to use this visualization to address a performance bottleneck.

%how a hardware designer can check on individual operations to understand the hardware and identify

%The file can be visualized in Chrome browser \url{chrome://tracing}. 
%\xxx[a]{This needs a little more explanation, and possibly a citation. Tell the reader what the Chrome tracing visualizer is, and why it's cool that you get to reuse it instead of making your own visualizer.}

\subsection{Simulation Engine} 
\label{sec:simulation-simulator}
%During simulation, the simulation engine keeps the status of each hardware component with a processor table to store the unique processor identifier, pointer to current operation, pointer to next operation and event queue storing all event operations. To specify the execution state of current operation, a processor table also keeps the time logs of the current operation it traces.

%the simulation engine also tracks the start time, end time and ready time of an operation.

The simulation engine loops over four stages: set up entry, check event queue, schedule operation, and finish operation.
The first stage sets up an \emph{operation entry} for each processor's current and next operation. % to prepare for the next three stages.
The second stage checks the head of each processor's event queue to decide whether to issue it. % the first operation on free processors.
The third stage models operations' execution time by updating the time logs in the operation entry.
To estimate execution time, the simulation engine uses a state object for each component.
For instance, a memory component uses banks, cycles per access, and read/write ports to calculate the time for a read or write operation.
Each component uses a \emph{schedule queue} to track operations and to model delay when contention happens, such as when two concurrent writes contend for the same memory.
The final stage models the effect of each operation by resetting its operation entry when it finishes.

\subsection{Extending the Simulator Library}

The simulation engine implements the primitives for EQueue programs using an extensible set of \emph{operation functions} and a \emph{component library}.
The interface for operation functions consists of a cycle count and a stall signal.
In the simulator's third stage, where it schedules operations (see previous section), it queries each operation function to obtain timing information.
At this point, the operation function may invoke a component object.

The EQueue infrastructure provides a standard library of components, such as SRAM memories and processors.
Designers can extend the library with custom components to introduce custom simulation logic.
To introduce a cache component, for example, a user would add a new \code{Cache} class to the component library and define an operation function to support \code{create_mem("Cache", ...)} in EQueue programs.
The operation function simply instantiates the cache component object.
The \code{Cache} class can inherit from a base \code{Memory} component class; the user only needs to override a method called \code{getReadOrWriteCycles} to determine whether the access is a hit or a miss and report a latency accordingly.
The \code{Memory} class inherits from a more general \code{Device} class that manages one or more scheduling queues to avoid conflicts.
In the case when there is conflict, the operation function returns a stall signal instead of a cycle count.
By extending these base classes, users can specify arbitrary behavior for components in EQueue programs.

\section{Lowering Passes}
\label{sec:passes}

The \equeue\ provides a set of reusable compiler passes.
Designers can combine these passes to build accelerators for simulation. We will show how to use these passes with the case study of systolic array in~\cref{sec:systolic-lowering}.

% \subsection{Component Field Specification Law} 
% Before we start, we introduce how to specify a component for EQueue lowering passes. A component field named \code{reg} inside a structure named \code{pe_array} can be denoted by \code{pe_array@reg}. %In the case \code{pe_array} is a vector, its indices come from the iterator of the closest affine loop. 
% One can also specify indices in a vector with subscription, e.g., \code{pe_array[5]@reg} denotes the \code{reg} field in the fifth component of \code{pe_array}. To allocate to one dimension of $N$-dimension vector, all dimensions except for specified dimension is represented by $[:]$, e.g., \code{pe_array[:][5]@reg} denotes \code{reg} field in the fifth column of \code{pe_array}.

% \subsection{Passes} 
\subsubsection{EQueue Read Write Pass} This pass translates \code{load} and \code{store} in MLIR's Affine dialect to EQueue's \code{read} and \code{write}.

\subsubsection{Allocate Memory Pass} This pass allocates buffers on a specified memory component.
% New buffers are added to the EQueue program's structure.
% The newly added buffer, e.g., \code{new_buffer} is added to the structure, e.g., \code{pe_array}. 
%In the future denotation, the component becomes \code{pe_array@new_buffer}.

\subsubsection{Launch Pass}
This pass adds \code{launch} operations by taking in a specified processor component and a code block.

\subsubsection{Memcpy Pass} This pass adds \code{memcpy} operations given specified source and destination buffer and a DMA component.

\subsubsection{Memcpy to Launch Pass} This pass changes a \code{memcpy} operation to \code{launch} with a block containing \code{read}s and \code{write}s.

\subsubsection{Split Launch Pass} This pass splits the specified launch block at the specified place.

\subsubsection{Merge Memcpy Launch Pass} This pass merges \code{memcpy} to the specified \code{launch} operation.
%by putting \code{read} to the front of the \code{launch} block and \code{write} to the end of the \code{launch} region. 
It avoids repetition if the \code{launch} block accesses the same buffer as the \code{memcpy}. %denotes the same as \code{read} or \code{write}. 
%and the same strategy applies to  destination buffer and \code{write} inside \code{launch} region.

\subsubsection{Reassign Buffer Pass} This pass replaces the uses of a buffer to another buffer. For instance, a SRAM read can be replaced with a register read. %in specified region. %For instance, a buffer read from \code{SRAM} can be updated with \code{pe_array[0][0]@reg} in a loop region.

\subsubsection{Parallel to EQueue Pass} This pass converts Affine dialect's \code{parallel} to  EQueue's \code{launch} with event dependencies.
%The EQueue events all depend on the same start event and use an \code{await} operation at the end to wait for all the events to finish.

\subsubsection{Lower Extraction Pass}
%\xxx[as]{I think this may need one extra sentence about what splat and extract do? Without that background, it's hard to see what structure is being crated here.}
This pass unrolls components denotation in vector form. 
%A component can be duplicated to form a vector with 
% \code{splat} in Standard dialect can duplicate a component to form a vector. This pass unrolls a \code{splat} to multiple \code{add_comp} and converts \code{extract} of Standard dialect to \code{get_comp} of \equeue.

\section{Case Study: Systolic array}
\label{sec:systolic}

\emph{Systolic array} is a widely-used mapping strategy to implement efficient multiplications and additions among matrices~\cite{kung1982systolic}.
As the communication is limited to neighbor processing elements (PEs), there is no cycle wasted on global communication and address matching. Because of their extremely broad design space of application-specific mapping strategies, memory systems, and PE designs, rapid simulation is critical to effectively exploiting systolic array designs.
%\xxx[a]{Suggested new motivation text: ``Because of their extremely broad design space of application-specific mapping strategies, memory systems, and PE designs, rapid simulation is critical to effectively exploiting systolic array designs.''}

In this section, we build and study an EQueue model of a systolic array.
We aim to answer these questions about the \equeue\ for this case study:
\begin{enumerate}
    \item Does embedding a simulator into a compiler framework help facilitate exploration of algorithmic mapping options? (\cref{sec:systolic-generator,sec:systolic-lowering})
    \item Can the simulation accurately estimate performance? (\cref{sec:systolic-accuracy})
    \item Is the simulation useful to help designers find the best design and does it scale? (\cref{sec:systolic-scalability})
\end{enumerate}
To answer question 1, we first show how to model a systolic array accelerator for convolutions using a \emph{generator} that emits a variety of configurations as EQueue programs.
We then also demonstrate a \emph{lowering pipeline} that translates from a high-level MLIR dialect into an EQueue model via a series of reusable compiler passes.
For question 2, we compare the EQueue simulation to a state-of-the-art custom simulator.
To address question 3, we measure our model to explore a design space of convolution accelerators.

% In this section, we introduce mapping different dataflows on systolic array and implement the mapping in two methods. The first method is to build a systolic array generator. The goal is to demonstrate the benefit of separation between dataflow modeling and simulation.
% %effectiveness of our simulator and explore the margin it can explore.
% The second method is to lower the same high-level abstraction with compiler passes to different dataflows, which avoids repeating work for hardware designers.
%with declarative builders API provided by MLIR \comment{cite?} 
%As dataflow mapping is a result of different tiling and structure matching strategy strategies, we show that properly aligned lowering passes on high level dialect can result in as good performance as generator-based systolic array.
%and prove our simulator is accurate in profiling cycles and bandwidth. We then learn the effect of array sizes and different dataflows on various convolution choices. Finally, a

\begin{figure*}[t!]
  \centering
\begin{subfigure}{0.3\textwidth}
  \centering
  \includegraphics[width=\textwidth]{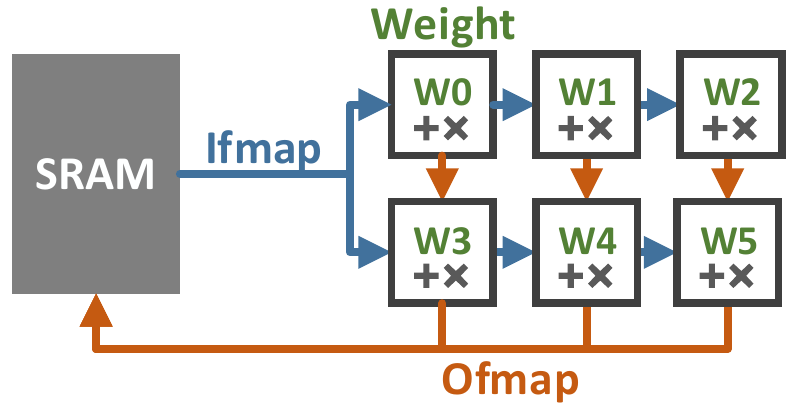}  
  \caption{Weight stationary.}
 \label{fig:dataflow-ws}
\end{subfigure}
\begin{subfigure}{0.3\textwidth}
  \centering
  \includegraphics[width=\textwidth]{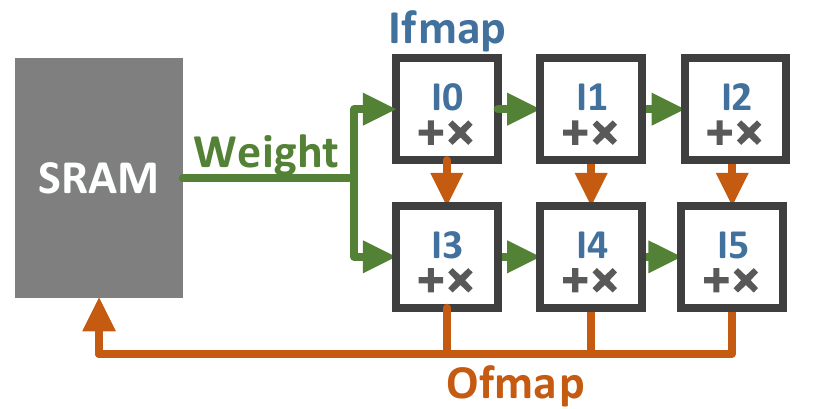}  
  \caption{Input stationary.}
 \label{fig:dataflow-is}
\end{subfigure}
\begin{subfigure}{0.3\textwidth}
  \centering
  \includegraphics[width=\textwidth]{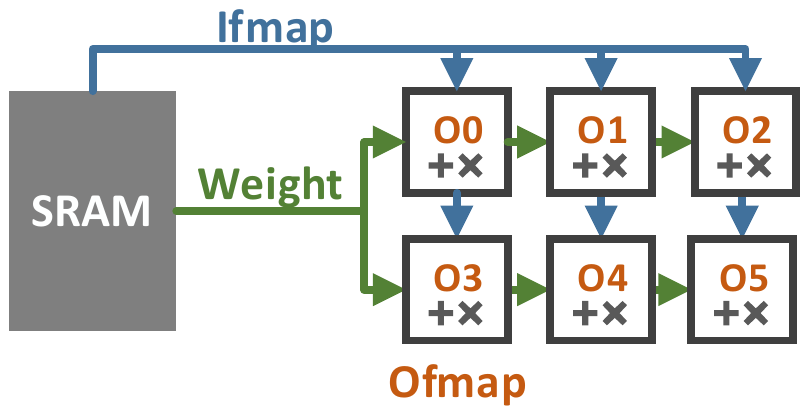}  
  \caption{Output stationary.}
 \label{fig:dataflow-os}
\end{subfigure}
\caption{Dataflows mapping on systolic array.}
\label{fig:dataflow}
\end{figure*}

\subsection{Background: Dataflows}
\label{sec:systolic-dataflow}

% A DNN accelerator with multiple processing elements (PEs) can perform computation in parallel.
%However, since nested loop computations like convolution can map several loops in spatial domain rather time domain, it leaves large design space on mapping strategies. 
A key design decision in a systolic accelerator implementation is the \emph{dataflow}, which determines how loops in the algorithm are mapped spatially onto processing elements (PEs)~\cite{sze2017efficient, chen2016eyeriss}.
% This section describes popular options for convolution dataflows in this case study.
In this case study, we consider
three widely-used dataflows:  Weight Stationary (WS), Input Stationary (IS), and Output Stationary (OS)~\cite{sze2017efficient}.
The difference is which tensor remains in each PE's register file:
the weights, input feature map (ifmap), or output feature map (ofmap).

Fig.~\ref{fig:dataflow} illustrates the data movement for each dataflow.
% A systolic array consists of identical PEs with fixed function.
On each cycle, each PE computes a part of final result and passes the partial result to its neighbor~\cite{lipton1985systolic}.
We use
$E_h, E_w$ for the ofmap height and width, $F_h, F_w$ for the filter height and width, $N$ for number of weights, and $C$ for channels.
Fig.~\ref{fig:dataflow-ws} shows that for WS, on each cycle, ifmaps and ofmaps are passed to the neighbor PEs, while each weight is stationary until $E_h \times E_w$ ifmaps convolve with it:
\begin{lstlisting}
pe[i+1][j].ofmap = pe[i][j].ofmap + pe[i][j].ifmap * pe[i][j].weight
pe[i][j+1].ifmap = pe[i][j].ifmap
\end{lstlisting} 
Fig.~\ref{fig:dataflow-is} shows IS.
On each cycle, weights and ofmaps are passed to the neighbor PEs, while every ifmap is stationary until $N$ weights convolve with it:
\begin{lstlisting}
pe[i+1][j].ofmap = pe[i][j].ofmap + pe[i][j].ifmap * pe[i][j].weight
pe[i][j+1].weight = pe[i][j].weight
\end{lstlisting}
Fig.~\ref{fig:dataflow-os} shows OS.
On each cycle, ifmaps and weights are passed to the neighbor PEs, while every ofmap is stationary until $F_h \times F_w \times C$ ifmaps and are convolved:
\begin{lstlisting}
pe[i][j].ofmap += pe[i][j].ifmap * pe[i][j].weight
pe[i+1][j].ifmap = pe[i][j].ifmap
pe[i][j+1].weight = pe[i][j].weight
\end{lstlisting}

% \begin{scriptsize}
% \begin{table}[t]
%   \centering
%   \caption{Experiment parameters.}
%   \label{table:exp-parameter}
%   \begin{tabular}{|l|l|l|l|}
%     \hline
%     \textbf{Parameters}&\textbf{Field} & \textbf{Representation} & \textbf{Value}\\
%     \hline
%     \hline
%     \multirow{2}{*}{PE Array }
%     &Array Height & $A_h$ & 2,4,8,16,32\\ \cline{2-4}
%     &Array Width & $A_w$ & $64/A_h$\\  \cline{2-4}
%     &Dataflow & - & IS/WS/OS\\ \hline
%     \multirow{6}{*}{Convolution }
%     &Ifmap Height & $H$ & 2,4,8,16,32\\ \cline{2-4}
%     &Ifmap Width & $W$ & $W$\\ \cline{2-4}
%     &Weight Height & $F_h$ & 1,2,4\\ \cline{2-4}
%     &Weight Width & $F_w$ & $F_h$\\ \cline{2-4}
%     &Ofmap Height & $E_h$ & $H-F_h+1$\\ \cline{2-4}
%     &Ofmap Width & $E_w$ & $W-F_w+1$\\\cline{2-4}
%     &Channels & $C$ & 1,2,4\\\cline{2-4}
%     &Weight Num & $N$&1,2,4,8,16,32\\\hline
    
%   \end{tabular}
% \end{table}
% \end{scriptsize}

\subsection{Systolic Array Generator}
\label{sec:systolic-generator}

This section demonstrates a \emph{generator} that emits EQueue code to model systolic array architectures.
We start with simple parallelism and build up to the full generator to illustrate the simplicity relative to a traditional, custom simulator.
% In this subsection, we introduce how to program a systolic array accelerator by gradually building an accelerator from simple parallelism to systolic passing and further introduce communication between different hierarchy to illustrate how our \equeue\ helps lower programmers' workload.
%, to introduce another level of communication from non-existence, while it can takes hours or days for experts to implement on a application-specific simulator. 

% for systolic array with declarative builders API of MLIR, i.e., writing sequential C++ programs to generate MLIR code. We also illustrate that benefiting from separation between architecture representation and simulation, the marginal effort to introduce another level of communication from non-existence, while it can takes hours or days for experts to implement on a application-specific simulator. 

%MLIR provides a declarative builders API, so that we can construct and manipulate MLIR programmatically, i.e., writing sequential C++ programs. We therefore implement a generator for systolic array generation in C++. 

\subsubsection{Parallelization}

We first show how to construct parallelism using \equeue.
We use MLIR's \emph{builder} API, which lets C++ code construct MLIR programs.
%which allows us to represent events and architectures in \equeue\ with \code{for} loops and array indexing.
This pseudo code shows a generator for a simple parallel architecture:
\begin{lstlisting}
start = control_start()
for h in arr_height:
  for w in arr_width:
    done = equeue.launch (...) 
      in (start, pe[h][w]){...} 
      // assume there is a PE array
    if w ==0 && h==0:
      prev_done = done
    else:
      prev_done = equeue.control_and(done, prev_done)
equeue.await(prev_done)
\end{lstlisting}
%
%We assume there is an array of processing elements (PEs) with  \code{arr_height} as height and \code{arr_width} as width. 
The \code{for} loop iterates over the dimension of the processing element (PE) array (\code{arr_height} by \code{arr_width}).
Each PE runs in parallel since they are all ``launched" by the same \code{control_start} event.
On each loop, \code{control_and} collects the \code{launch} events of the current and previous PE.
An \code{await} barrier ensures that the current processor waits for all \code{launch} events to finish.
We later denote this pattern as \code{par_for}.

\subsubsection{Systolic passing}

We next extend the generator to pass values systolically between PEs.
We use two stages:
one reads values from a buffer and compute results, and a second stage passes values to neighboring PEs.
This generator code shows WS dataflow and omits boundary conditions for simplicity:
%\xxx[as]{I am a little concerned that this code block is just too long to follow... it may be unfixable now, but I wish there were a way to simplify it to highlight exactly what readers should be paying attention to.}
%
\begin{lstlisting}
//reading stage
par_for (h, w) in arr_height-1, arr_width-1:
  done, weight_value[h][w], ofmap_value[h][w]=
    equeue.launch(
    weight_buffer = (*@\textcolor{red}{pe[h][w].weight\_buffer}@*),
    ifmap_buffer = (*@\textcolor{red}{pe[h][w].ifmap\_buffer}@*),
    ofmap_buffer = (*@\textcolor{red}{pe[h][w].ofmap\_buffer}@*))
    in (start, pe[h][w].kernel) {
    ifmap = equeue.read(ifmap_buffer)
    weight = equeue.read(weight_buffer)
    ofmap_old = equeue.read(ofmap_buffer)
    ofmap = ifmap * weight + ofmap_old
    equeue.return weight, ofmap
  }
//writing stage
par_for (h, w) in 1 to arr_height, 1 to arr_width:
  done = equeue.launch(
    weight = weight_value[h][w],
    ofmap = ofmap_value[h][w],
    weight_buffer = (*@\textcolor{red}{pe[h][w+1].weight\_buffer}@*),
    ofmap_buffer = (*@\textcolor{red}{pe[h+1][w].ofmap\_buffer}@*))
    in (start, pe[h][w].kernel) {
    equeue.write(ofmap, ofmap_buffer)
    equeue.write(weight, weight_buffer)
  }
\end{lstlisting}
In the read stage, each PE reads
ifmap, weight and ofmap values from corresponding buffers and computes an ofmap.
In the write stage, the PEs in each column (\code{pe[h][w]}) pass weights to the next column (\code{pe[h][w+1]}).
PEs in a given row write ofmaps to buffers in the next row (\code{pe[h+1][w]}).
%\xxx[as]{Please double-check that this is still right; I simplified it a lot.}

\subsubsection{Model SRAM Bandwidth}
So far, we have a complete and cycle-accurate model of the core PE array logic.
The next step is to model the PE array's interaction with associated SRAMs to measure read and write bandwidth.
Extending our EQueue generator, we can change the \code{launch} input in our read stage to
make the first column of PEs read from an SRAM:
\begin{lstlisting}
par_for (h, w) in arr_height, arr_width:
  if w == 0: buffer = sram.ifmap_buffer
  else: buffer = pe[h][w].ifmap_buffer
  done, ... = equeue.launch( ifmap_buffer = buffer,
    ...){ ... } // other code same as before
\end{lstlisting}
%
% The above program shows the modification to the reading stage of systolic passing. The first column of the PE array reads ifmaps from buffers on SRAM rather than buffers on PEs. %in the second stage, the ifmaps are still write to PE buffers.
Similarly, we can modify the write stage to store ofmaps from the last row of PEs to an SRAM:
\begin{lstlisting}
par_for (h, w) in arr_height, arr_width:
  if w == arr_height-1: obuffer=sram.ofmap_buffer
  else: obuffer=pe[h+1][w].ofmap_buffer
  done = equeue.launch( ofmap_buffer = obuffer,
    ...) {...} // other code same as before
\end{lstlisting}
With these small changes,
the simulation engine can model communication between SRAMs and the PE array.

\noindent \textbf{Benefits.}
EQueue programs can modularize hardware components (e.g., SRAM interfaces and processors) and thereby study the individual effect of a component.
Separating representation from simulation allows a programmer to concentrate on architecture design; no changes are necessary to the simulation engine to evolve the modeled hardware.

% This example shows how EQueue generators can gradually evolve to model more parts of the target architecture.
% Separating representation from simulation allows a programmer to concentrate on architecture design; no changes are necessary to the simulation engine.
% Adding bandwidth sensitivity to a custom C++ simulator, in contrast, would require manual changes to the simulation strategy.

\subsection{Comparison with SCALE-Sim}
\label{sec:systolic-accuracy}

\begin{figure*}[ht]
\begin{subfigure}[t]{.24\textwidth}
  \centering
  \includegraphics[width=\linewidth]{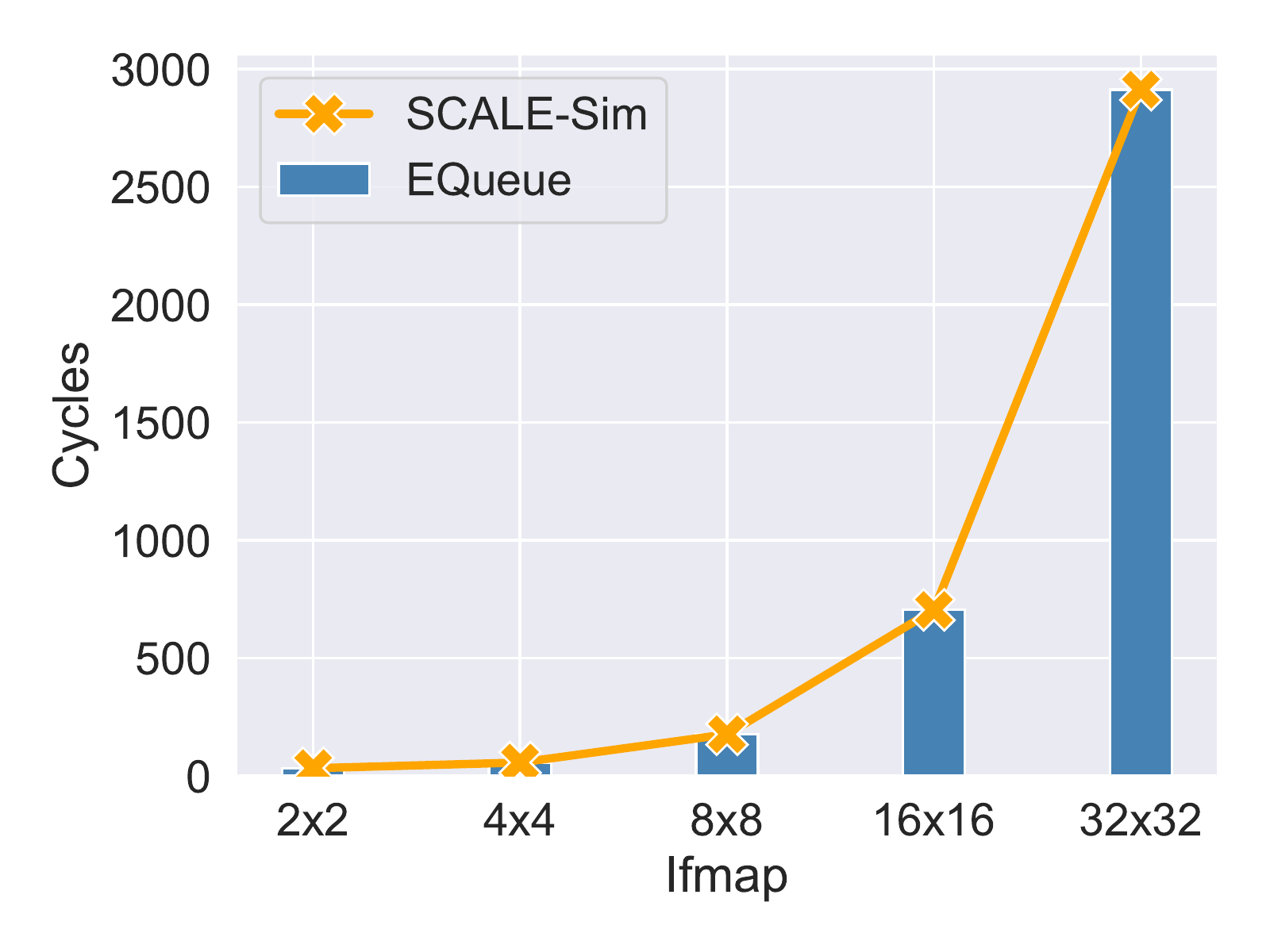}  
  \caption{Cycles.}
 \label{fig:cycle_ifmap}
\end{subfigure}
\begin{subfigure}[t]{.24\textwidth}
  \centering
  \includegraphics[width=\linewidth]{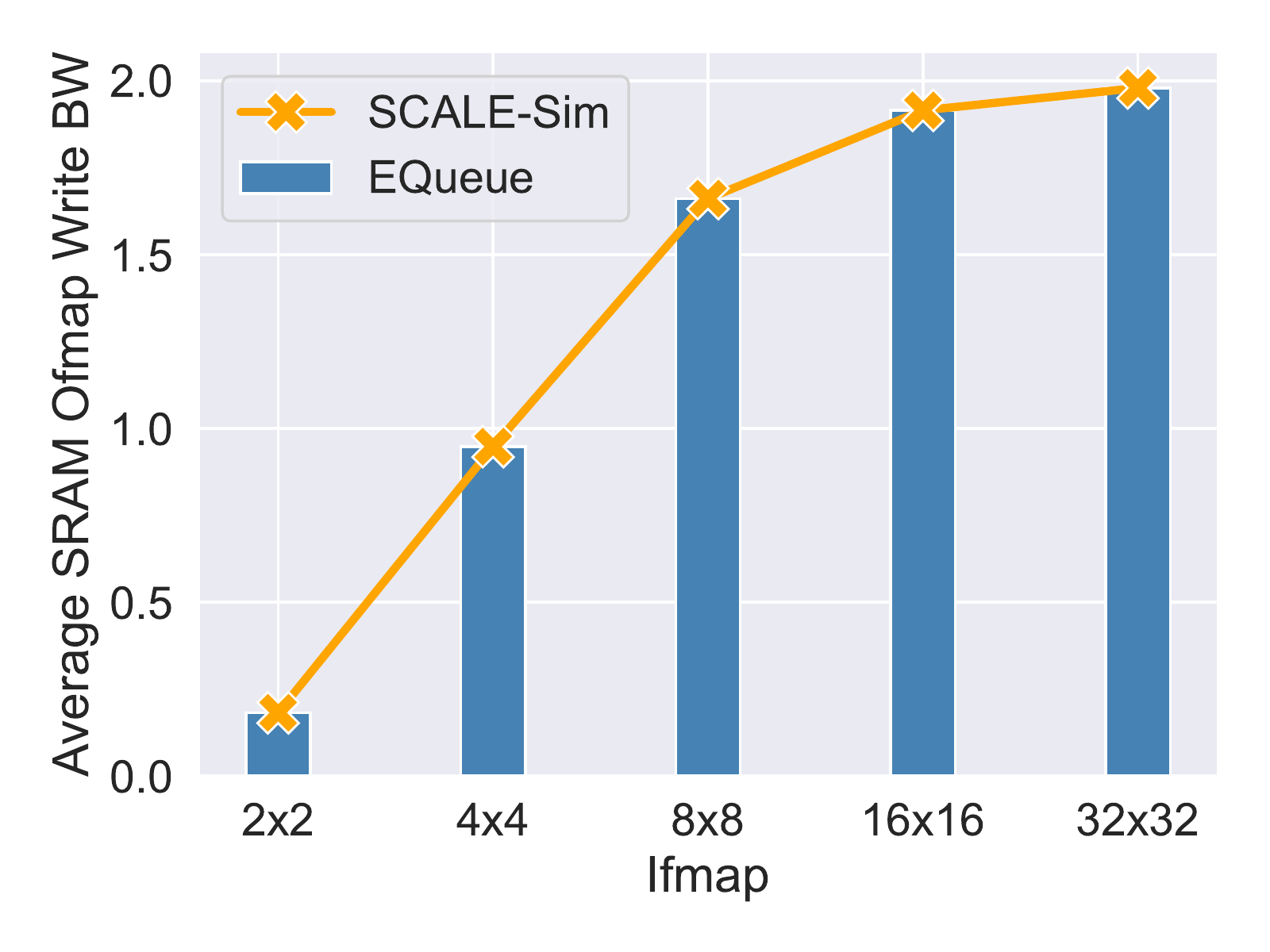}  
  \caption{SRAM write BW.}
 \label{fig:bw_ifmap}
\end{subfigure}
\begin{subfigure}[t]{.24\textwidth}
  \centering
  \includegraphics[width=\linewidth]{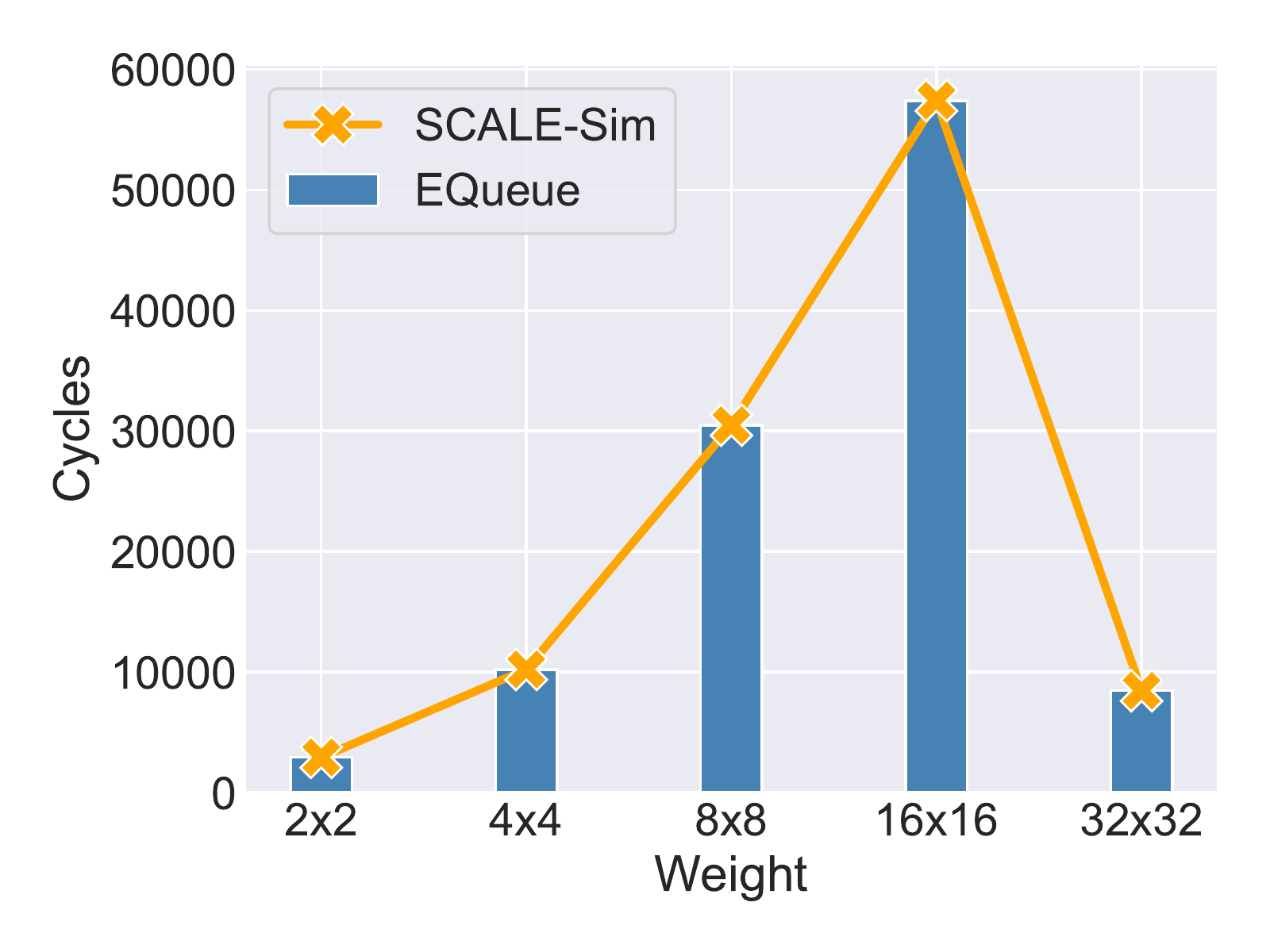}  
  \caption{Cycles.}
 \label{fig:cycle_weight}
\end{subfigure}
\begin{subfigure}[t]{.24\textwidth}
  \centering
  \includegraphics[width=\linewidth]{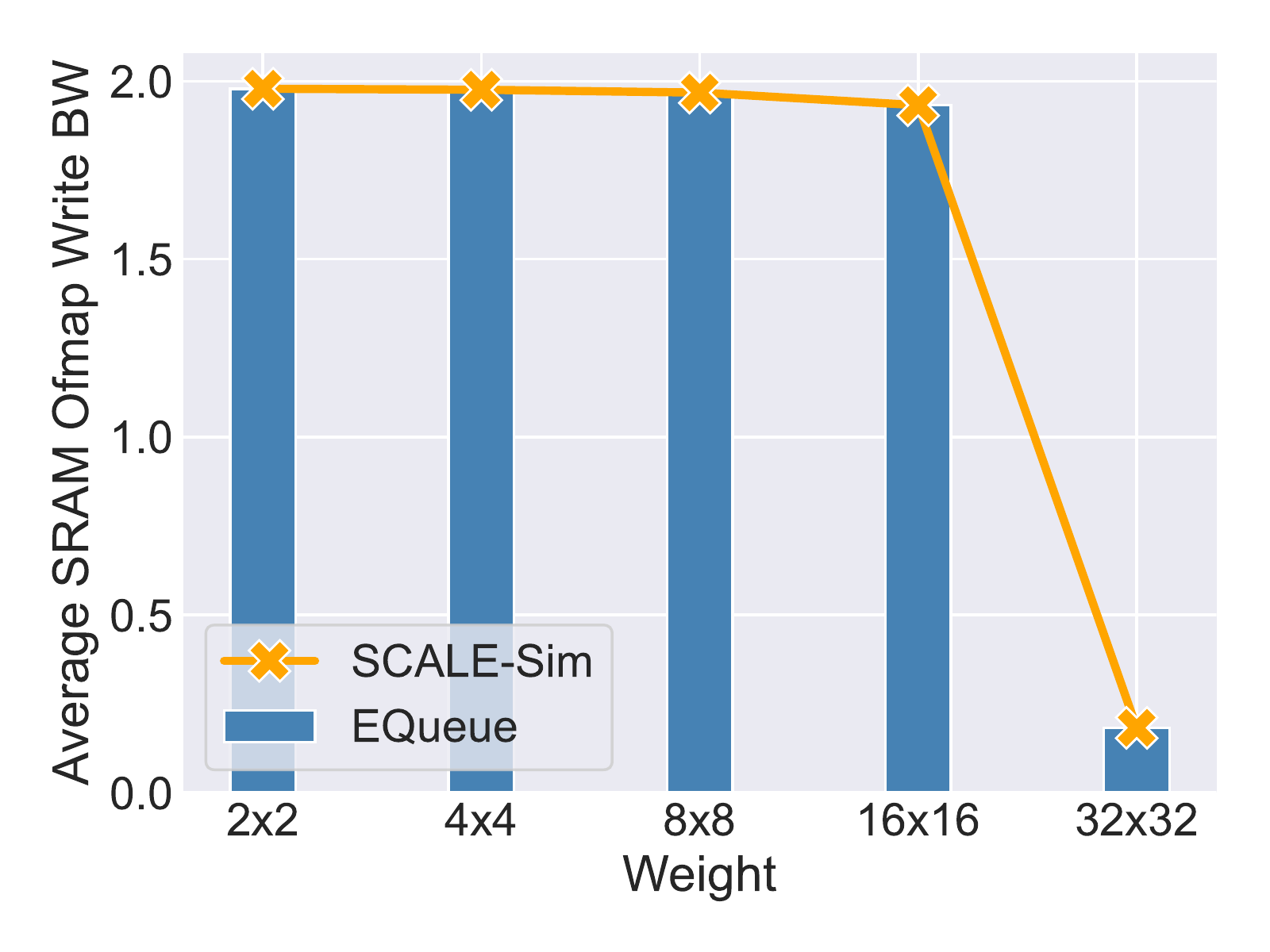}  
  \caption{SRAM write BW.}
 \label{fig:bw_weight}
\end{subfigure}
\caption{Comparing our simulation with SCALE-Sim on 4$\times$4 systolic array, by convolving various ifmaps with fixed $2\times2\times3$ weights (a--b) and by convolving various weights convolved with fixed $32\times32$ weights (c--d).}
\label{fig:compare-to-scalesim}
\end{figure*}

To check the accuracy of our systolic array EQueue model, we compare to a validated simulator
SCALE-Sim~\cite{samajdar2018scale} specific to WS, IS, and OS convolutions on systolic array. %using three separate implementations.

Fig.~\ref{fig:compare-to-scalesim} compares the simulated cycles and average bandwidth for our model and SCALE-Sim, both modeling a 4$\times$4 WS systolic array with various ifmap and weight sizes.
Our EQueue-based simulation matches SCALE-Sim's results.

\noindent \textbf{Benefits and Costs.}
When exploring design alternatives, an EQueue-based simulator has a lower programming cost than a custom one-off simulator.
SCALE-Sim~\cite{samajdar2018scale}'s WS and IS implementation have little code overlap: WS is implemented in Python in 569 lines of code (LOC),
%and IS is implemented in 500 LOC
but switching from WS to IS requires changing 410 LOC.
In contrast, our EQueue program for WS is implemented in C++ in 281 LOC only needs 11 LOC to switch from WS to IS.

In exchange, the one-off simulator has a performance advantage: for experiments in Fig.~\ref{fig:compare-to-scalesim}, SCALE-Sim takes at most 1.1 second, while the EQueue simulator takes at most 7.2 seconds.
The speed comes at the cost of complex modifications while exploring the architectures and algorithm mappings.

\subsection{Lowering Pipeline}
\label{sec:systolic-lowering}

\begin{figure*}[t]
  \centering
  \includegraphics[width=0.95\linewidth]{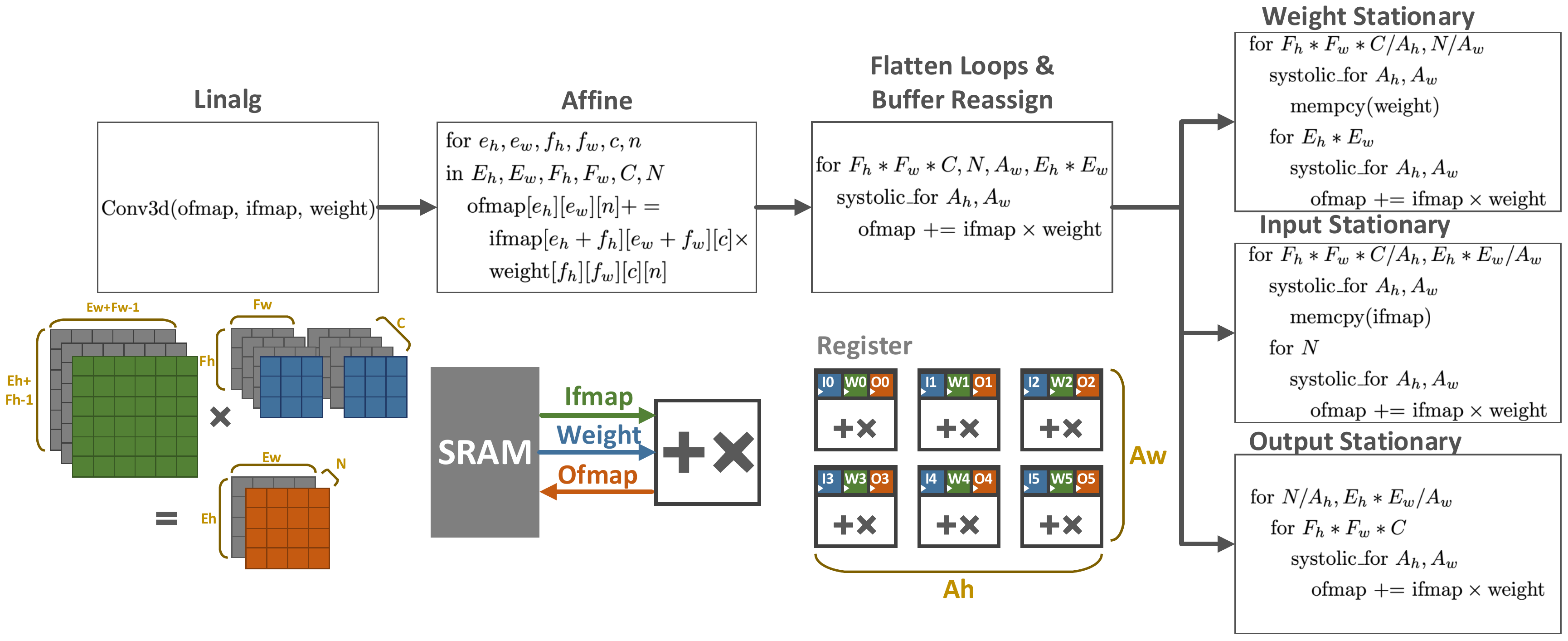}  
  \caption{Lowering pipeline for WS, IS, OS dataflow. They share the same lowering stages except the last one.}
 \label{fig:lowering-pipeline}

\end{figure*}

%In this section, we discuss how to lower convolution to specific dataflow by applying our lowering passes.

%While direct hardware model generation may be appraise when the target audience is an expert, the \equeue\ also makes it easy for software and algorithm developers to generate custom simulators.
%So far we have shown that separation between hardware and simulation lowers the programming overhead while providing as accurate and useful results as traditional simulators, 

% A main benefit of compiler-driven simulation is it can combine reusable compiler passes to lower and transform programs.
\noindent \textbf{Rationale.} The benefit of a compiler-driven approach is not limited to lowering the bar of programming:
more importantly, it makes it possible to program the simulator using compiler passes.
Integrating with a compiler stack's shared passes avoids the need for tedious, manual modification to explore different program mappings.

This section constructs a \emph{lowering pipeline} that compiles from high-level algorithmic specifications to EQueue hardware models.
Critically, the pipeline can produce different dataflows for the same input program by applying different sequences of compiler passes.

% The benefit of a compiler-driven approach is not limited to lowering the bar of programming, but more importantly, implementing different designs with minimal cost by sharing compiler passes. 
% This section shows how to implement systolic array accelerators of different dataflows with a lowering pipeline of common passes, applied via different orders and parameters.
 %to the same architecture. Linalg dialect, a high level dialect for tensor computation.

%Fig.~\ref{fig:dataflow} shows the common ground between the three different dataflows: at each cycle the value from one PE is read, modified and written to its neighbor PE. Therefore, one can construct a lowering pipeline sharing many intermediate stages and passes.

%The opposite strategy is to lower from the same high-level dialect and share some common lowering stages. Even when the three dataflows have to diverge at certain stage, they can still use the same passes but width different parameters and order to apply. 

%reusable lowering passes facilitate common transformations and 
%To try a dataflow, designers can apply different arrangement of passes and pass parameters, rather than implementing representation or even simulator from the beginning.% as most of simulators require. 
%share the passes during lowering and separate mapping strategies with a proper arrangement of passes and pass parameters. 

\smallskip
\noindent
\textbf{Implementation.}
IS, WS, and OS all share a core systolic design: on each cycle, each PE reads a value, modify it, and writes to a neighbor PE.
Fig.~\ref{fig:lowering-pipeline} shows how the systolic dataflows share stages along a lowering pipeline.
The first 3 stages (Linalg, Affine, and Reassign) are the same.
The final stage (Systolic) diverges, but lowering from Reassign stage allows different dataflows to share lowering passes with different orders and parameters. This way, hardware designers can only implement the highest level abstraction and then explore design spaces with no programming overhead.
%explains the steps to map the convolution represented by Linalg dialect to a specific dataflow represented by \equeue. 

\subsubsection{Linalg to Affine}

We start with a convolution in Linalg dialect, an MLIR dialect that can express arbitrary linear algebra.
The Linalg dialect can be first lowered to the Affine dialect with the standard \code{--convert-linalg-to-affine-loops}, which lowers the convolution to explicit nested loops.
We then apply \code{--equeue-read-write} to change \code{load} and \code{store} operations in Affine dialect to \code{read} and \code{write} in \equeue\ to model data movement.

\subsubsection{Affine to Buffer Reassign}

Next, we apply the \code{--allocate-buffer} and \code{--reassign-buffer} passes to replace direct SRAM reads and writes with PE local register accesses.
At this stage, we also flatten the 6 convolutional dimensions ($E_h, E_w$, $N$, $F_h, F_w, C$)
into 3: $E_h \times E_w$, $N$, $F_h \times F_w*C$.
This flattening reflects the stationary dimension on PEs for each dataflow:
for WS, each weight is stationary on a PE until computed with $E_h \times E_w$ ifmaps; for IS, each ifmap is stationary for $N$ weights; for OS, each ofmap is stationary until accumulated with $F_h \times F_w \times C$ ifmaps and weights.

\subsubsection{Buffer Reassign to Systolic Array}

After flattening, for WS and IS, we first need to copy weights or ifmaps from the SRAM into the PE array registers.
We generate the necessary \code{memcpy} operations with a
\code{--mem-copy} pass and merge them with \code{launch} operations using \code{--merge-memcpy-launch}.
Then, we implement systolic communication.
% Let $A_h$ and $A_W$ denote the PE array's height and width.
For WS, we need to pass the ifmaps and ofmaps to the right and down on every cycle.
% are passed in a systolic way, i.e., at each cycle, $A_h-1$ ifmaps and $A_w-1$ ofmaps are read from local PE passed to its right and down.
Similarly, for IS, $N$ weights and ofmaps are passed, while for OS, $F_h \times F_w*C$ ifmaps and weights are passed.
The \code{--split-launch} and \code{--reassign-buffer} passes implement this systolic communication.
% The systolic style of communication from each PE to its neighbor can be implemented with \code{--split-launch} pass and \code{--reassign-buffer} pass. %on the splitted block by reassigning buffers at one column of PE array to the right. %\code{pe_array[:]@pe_wbuffer} to \code{pe_array[+1]@pe_wbuffer}.\xxx[zl]{not sure how about representation}.
Finally, we apply \code{--parallel-to-equeue} and \code{--lower-extraction} passes to complete lower operations to \equeue.

The key advantage of the lowering pipeline approach is the reduced effort for implementing different dataflows.
In a traditional simulator, changing the mapping strategy requires extensive rewriting of the simulation engine.
In the compiler-driven approach, designers can apply different combinations of reusable passes to try out different dataflows.
%To facilitate simulation, we also apply \code{--parallel-to-equeue} pass to lower parallel operation to equeue launching operations triggered by the same start signal, generating done signals to the same await operation, and \code{--lower-extraction} pass to lower PE array represented by vector to PEs with different offset in names.

\begin{figure*}[t]
\begin{subfigure}[t]{.245\textwidth}
  \centering
  \includegraphics[width=\linewidth]{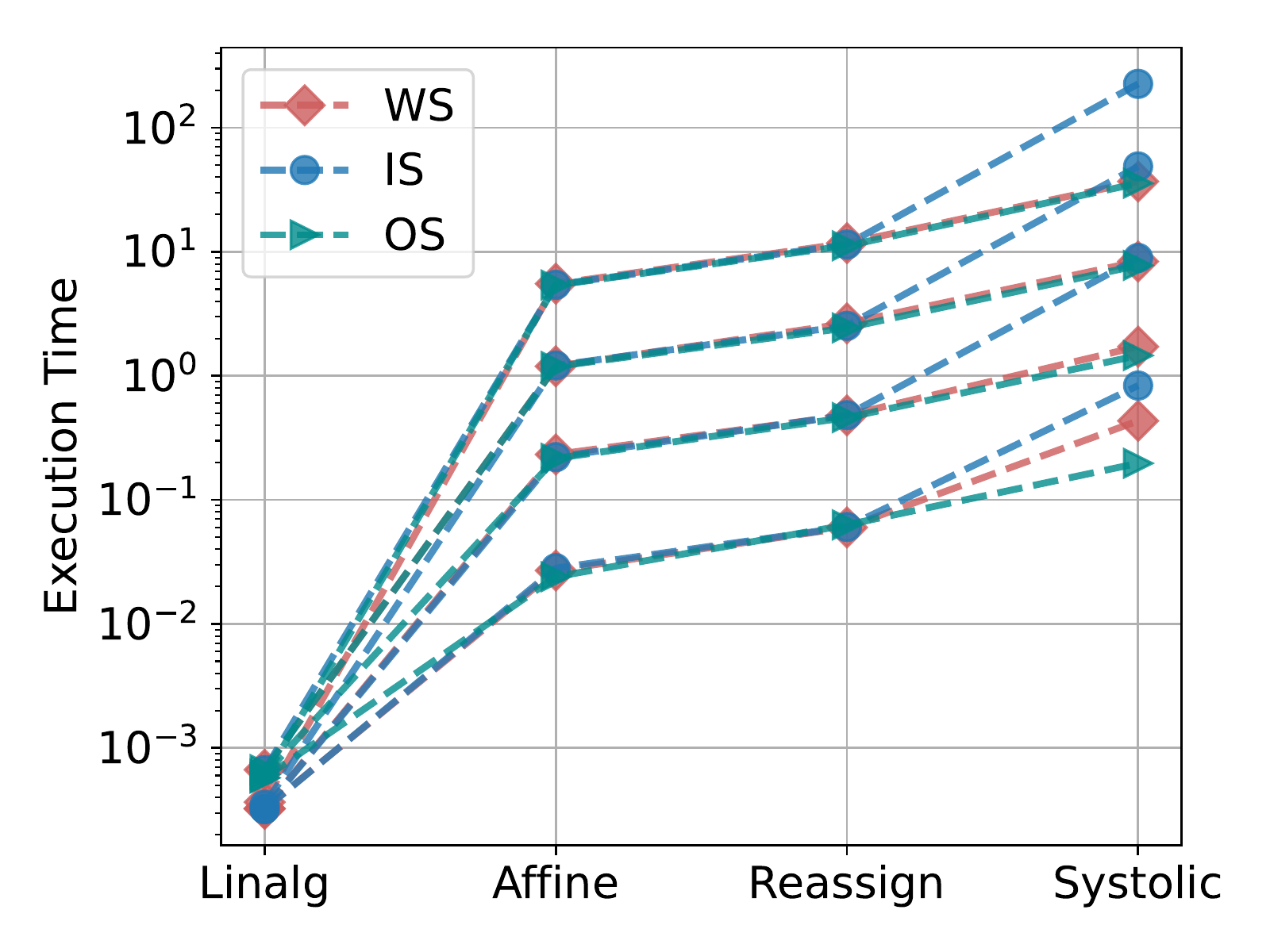}  
  \caption{Execution time for simulation.}
 \label{fig:lowering-exec-time}
\end{subfigure}
\begin{subfigure}[t]{.245\textwidth}
  \centering
  \includegraphics[width=\linewidth]{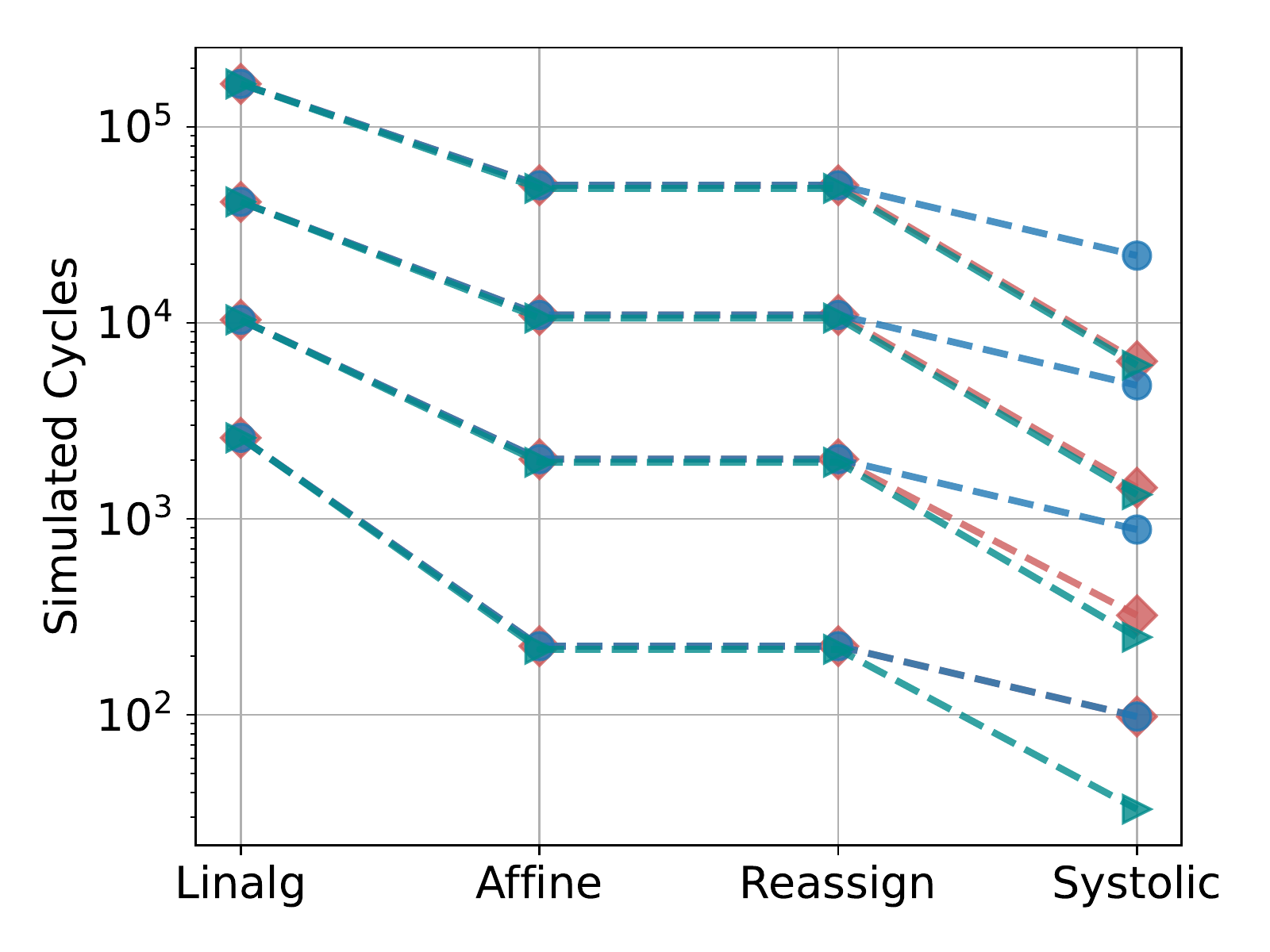}  
  \caption{Runtime in cycles.}
 \label{fig:lowering-cycles}
\end{subfigure}
\begin{subfigure}[t]{.245\textwidth}
  \centering
  \includegraphics[width=\linewidth]{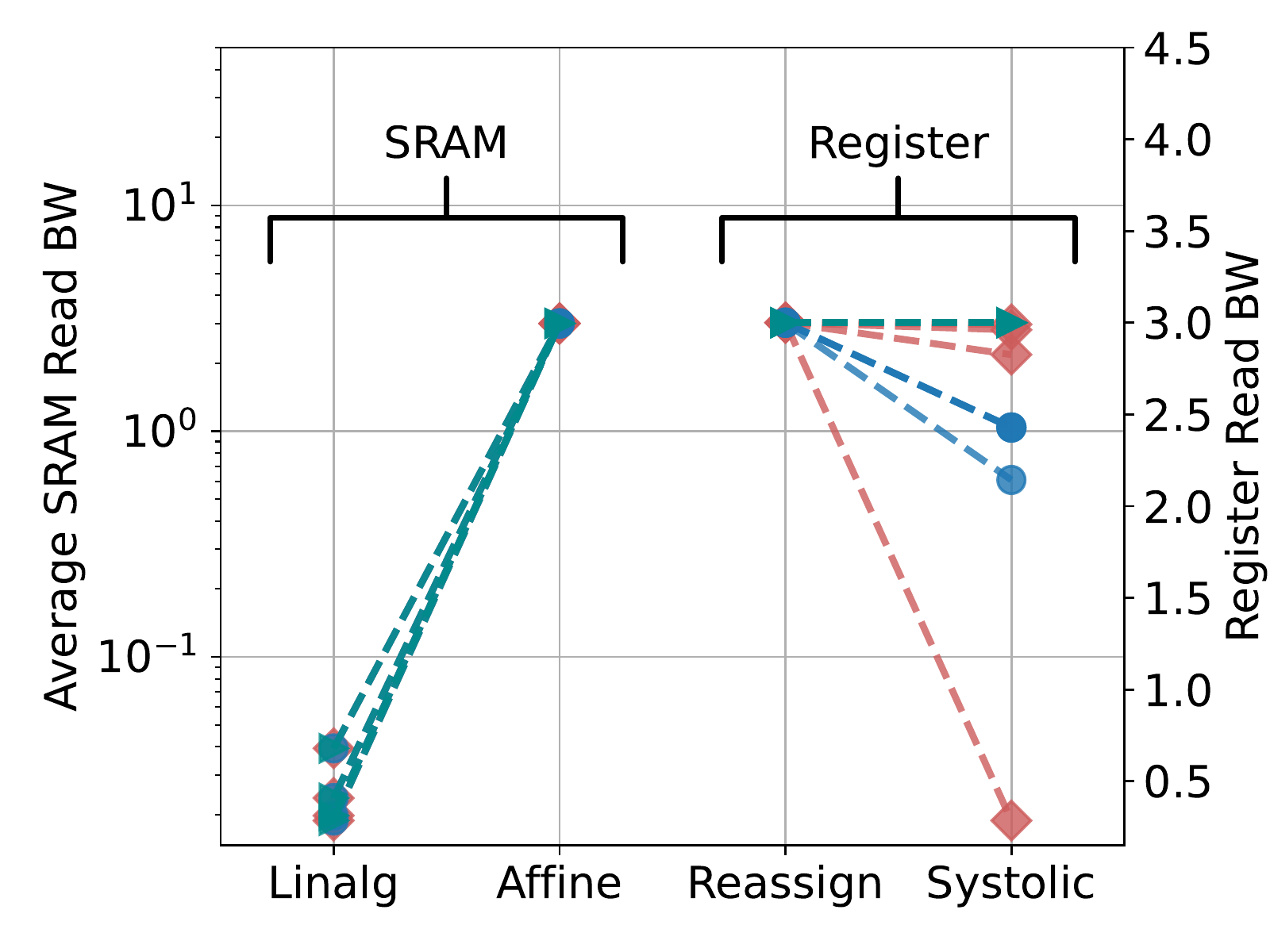}  
  \caption{Read bandwidth.}
 \label{fig:lowering-read-bw}
\end{subfigure}
\begin{subfigure}[t]{.245\textwidth}
  \centering
  \includegraphics[width=\linewidth]{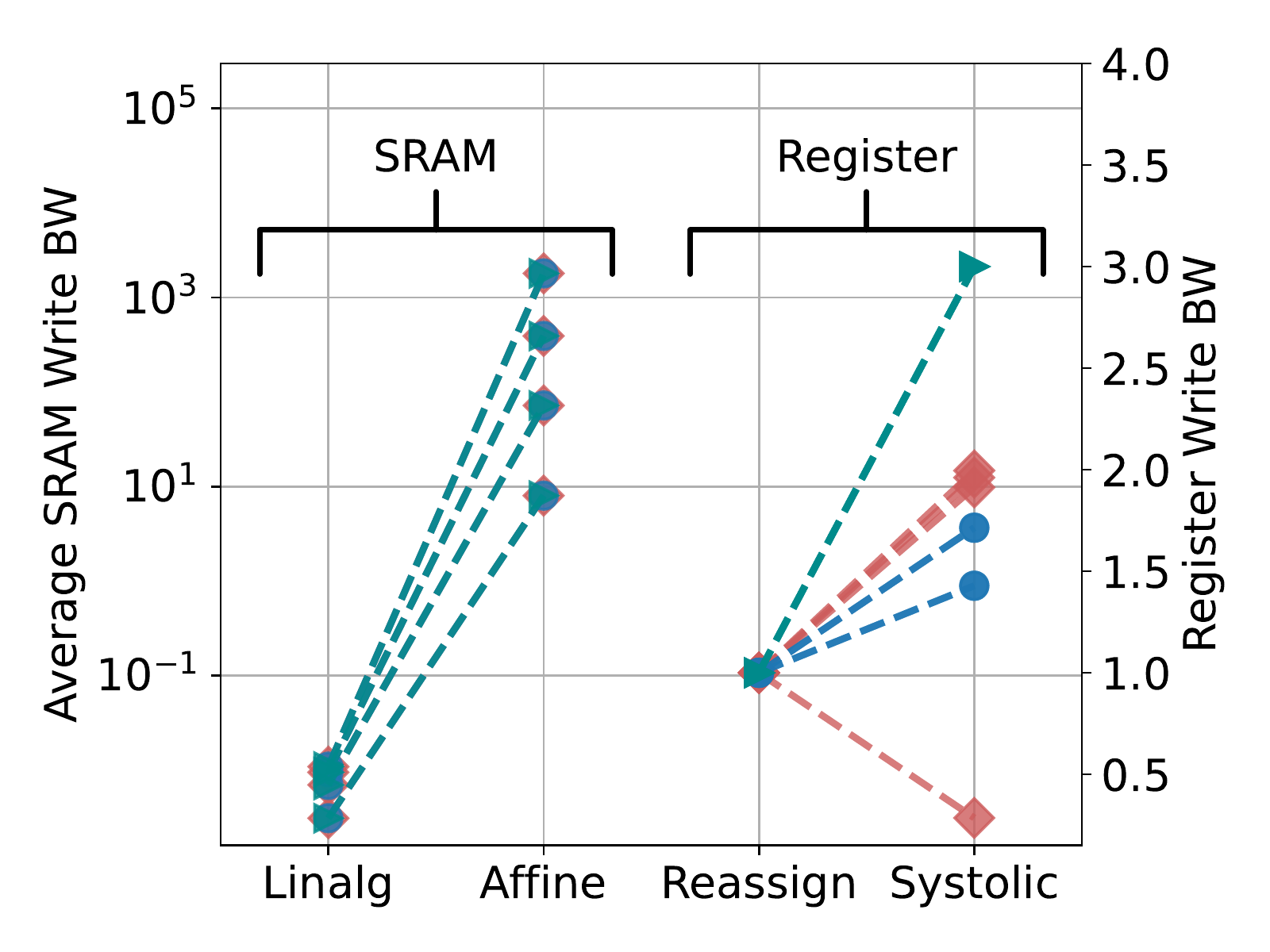}  
  \caption{Write bandwidth.}
 \label{fig:lowering-write-bw}
\end{subfigure}
\caption{Figure showing various metrics along the four stages of lowering pipeline on an accelerator with a 4$\times$4 PE array and convolution settings $H=W=4,8,16,32, F_h=F_w=3, C=3, N=4$. Compilation time is negligible and omitted.}
\label{fig:lowering-results}
\end{figure*}

\smallskip
\noindent
\textbf{Results.}
% The EQueue framework can simulate hardware at multiple levels of abstraction.
% We next study this capability by measuring the simulated hardware at different points along the lowering pipeline.
Fig.~\ref{fig:lowering-results} plots the simulator execution time, simulated runtime, and read and write bandwidth on the four convolution settings at the four stages (Linalg, Affine, Reassign, and Systolic).
This compiler pipeline does not take a significant amount of time (it typically finishes in microseconds).

The first three lowering stages are identical for different dataflows, so
they have the same bandwidth and runtime.
This sharing reflects the framework's reusability for common parts of different accelerator implementations.
At the final stage, the runtime differs from the simpler generator-based approach from Section~\ref{sec:systolic-generator} by 1.2\% on average, up to 2\%.
The difference lies in warm-up and cool-down phases that the passes do not model.
Register and SRAM bandwidth differs for the same reason.

% To learn the accuracy of the lowering pipeline, 
% %To learn if the systolic array with lowering pipeline implementation can provide accurate simulation, 
% we compare the runtime in cycle at Systolic stage with generator-based approach on the four convolution settinngs (H=W=4,8,16,32).

%\xxx[zl]{not sure if we want this \& where to put}
Fig.~\ref{fig:lowering-results} also reflects the transition of hardware at each stage.
From Linalg stage to Affine stage, the execution time grows, the runtime reduces, and the SRAM bandwidths grow, since affine stage models explicit nested loops and data movements.
%\xxx[as]{This is an extremely long sentence. Can we break it up into smaller, more direct sentences?}
%The Affine stage models explicit nested loops and data movements; 
At the Reassign stage, we model reads and writes on registers rather than SRAM, so the register bandwidth changes from 0 to 1 byte per cycle and the execution time grows. 
%Only one processor does the computation and the runtime does not change.
At systolic stage, we introduce a grid of PEs running concurrently, resulting in higher execution time, lower runtime in cycles and differentiated bandwidth.
 
%as the fir
%because they share the same representation designers as they only need to adjust the passes to lower to the last stage to obtain different dataflow. 

% The experiment shows the effectiveness of compiler passes to implement different designs, as other simulators require rewriting the simulation inputs or simulators.
% As the workflow changes, a designer only needs to change the lowering pass order and parameters to achieve different mappings.

\noindent \textbf{Benefits.}
The availability of reusable lowering passes lets designers rapidly switch between program--accelerator mappings and
enables efficient design space exploration.
In contrast, one-off simulators would require custom modifications to support these transformations.%, while general-purpose simulation infrastructures like gem5~\cite{binkert2011gem5} would require reimplementing the architecture.
\xxx[as]{While the first sentence here is compelling, I'm not 100\% convinced about the gem5 notion. General-purpose simulators would have a hard time accomplishing this simulation in the first place, so I'm not sure it's a useful comparison. Maybe we should stick with just the first part of the criticism, which is easier to agree with?}

%The evaluation demonstrates lowering pipeline can provide accurate results without manually writing a generator for each of the dataflow.

%the substantial time and efforts the compiler-based simulation can save for hardware designers.%, while it leaves a lot of spaces for further improvement by extending MLIR framework. 

%For hardware designers, it saves a lot of efforts to implement similar designs when a toolchain can provide common transition stages and reusable passes. 
% We evaluate our lowering pipeline on systolic array proposed in \cref{sec:systolic-lowering}. There are several questions we want to anser:
% \textit{
% \begin{enumerate}
%     \item Is the pipeline implemented systolic array accurate?
%     \item What is the indication of profiling results along the pipeline?
% \end{enumerate}
% }
%It is essential that the lowering pipeline is capable of providing comparable results as generator. 

% For a systolic array, WS, IS and OS can easily share the common transition stages and reuse most of the passes, e.g., buffer reassigning and splitting launch regions, in the final stages. For an algorithm designer, the time to try different dataflows is hugely economized as compiling different dataflow takes micro seconds. 

\subsection{Scalability Evaluation}
\label{sec:systolic-scalability}

\begin{figure*}[t]
%\begin{minipage}{0.39\textwidth}

\begin{subfigure}[t]{.19\textwidth}
  \centering
  \includegraphics[width=\linewidth]{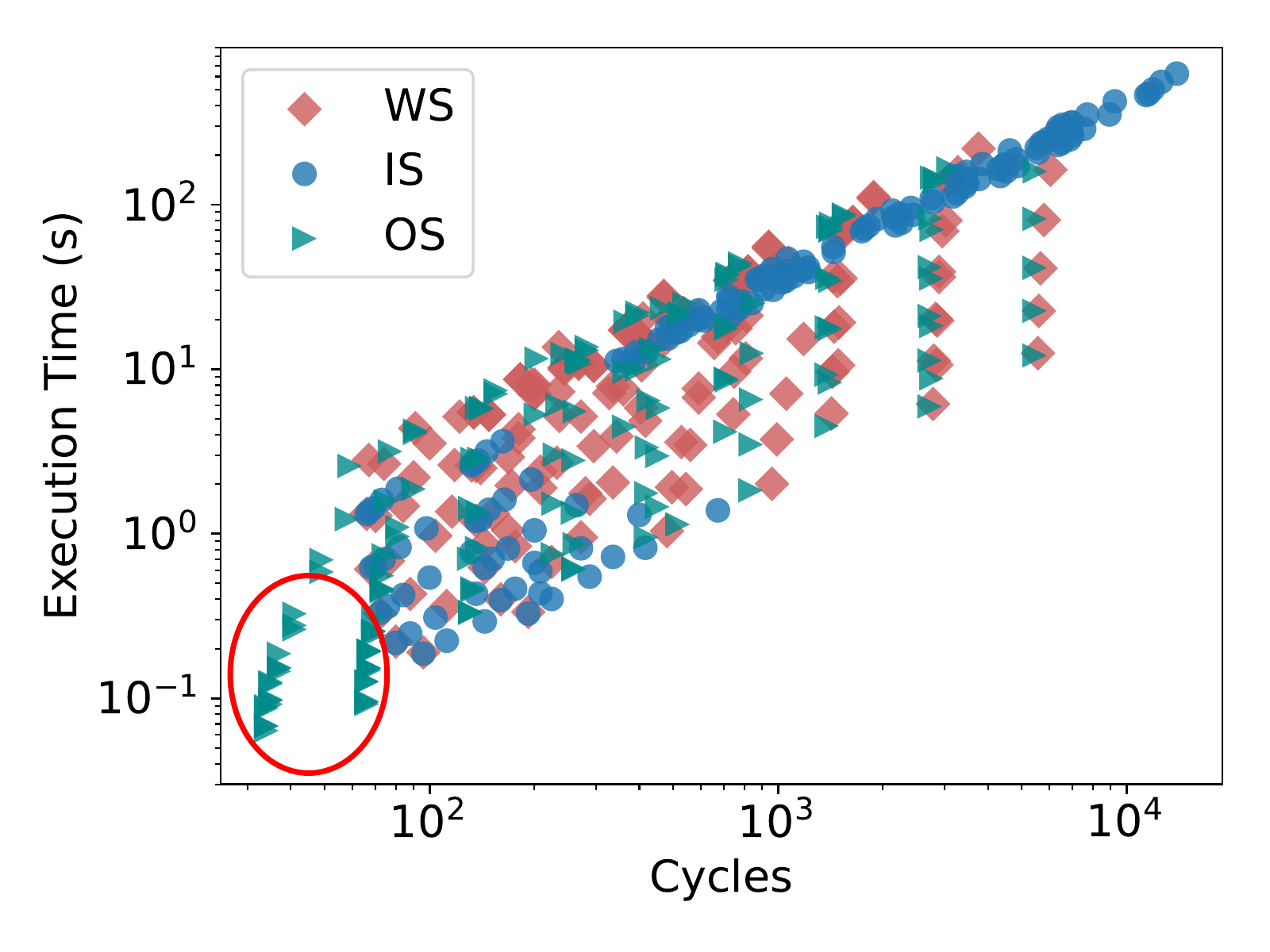}
  \caption{Execution time.}
  \label{fig:cycle-exec-time}
\end{subfigure}
\begin{subfigure}[t]{.19\textwidth}
  \centering
  \includegraphics[width=\linewidth]{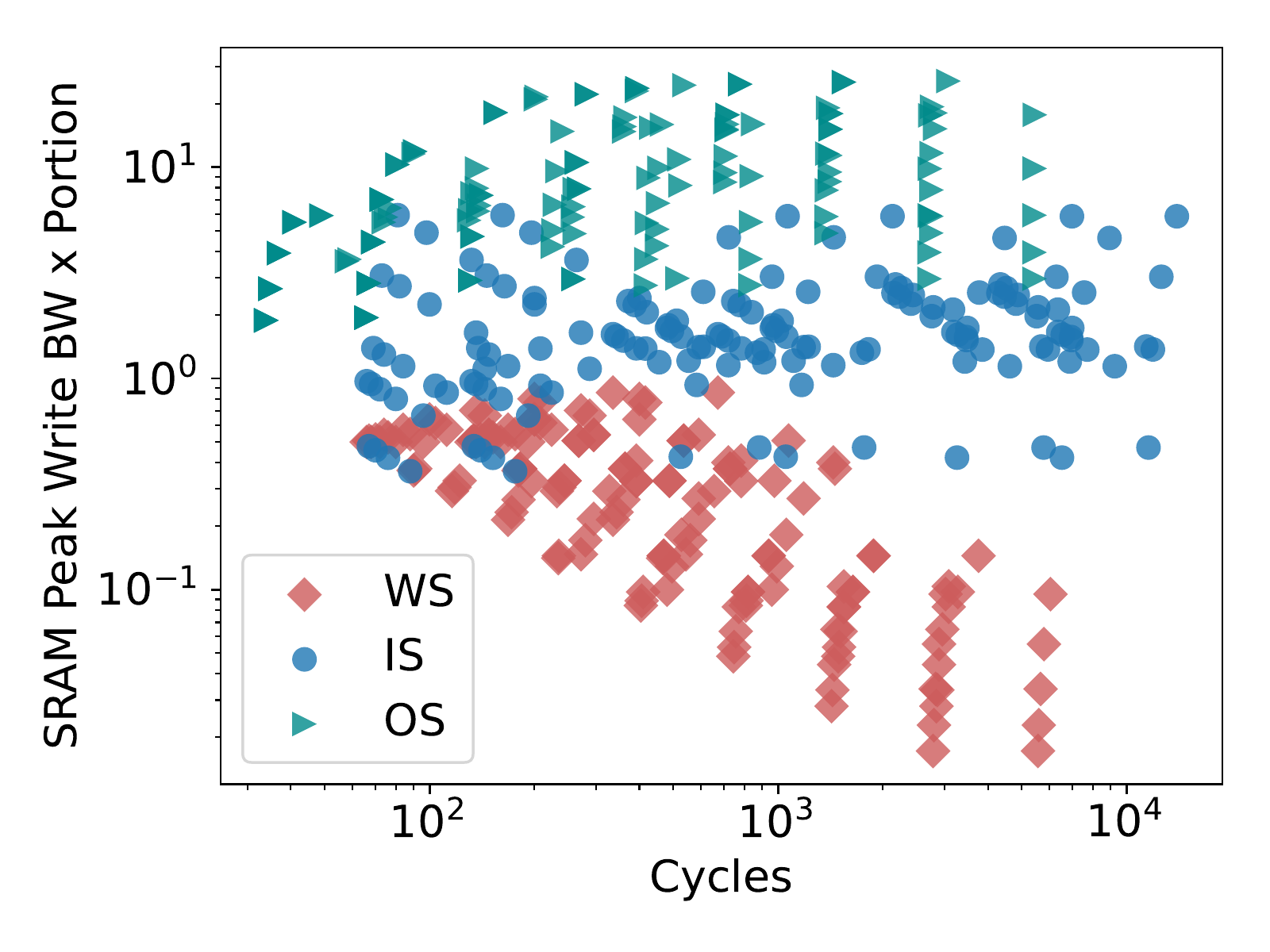}
  \caption{Peak read bandwidth.}
  \label{fig:cycle-bw-read}
\end{subfigure}
%\end{minipage}\hfill
%\begin{minipage}{0.59\textwidth}
\begin{subfigure}[t]{.19\textwidth}
  \centering
  \includegraphics[width=\linewidth]{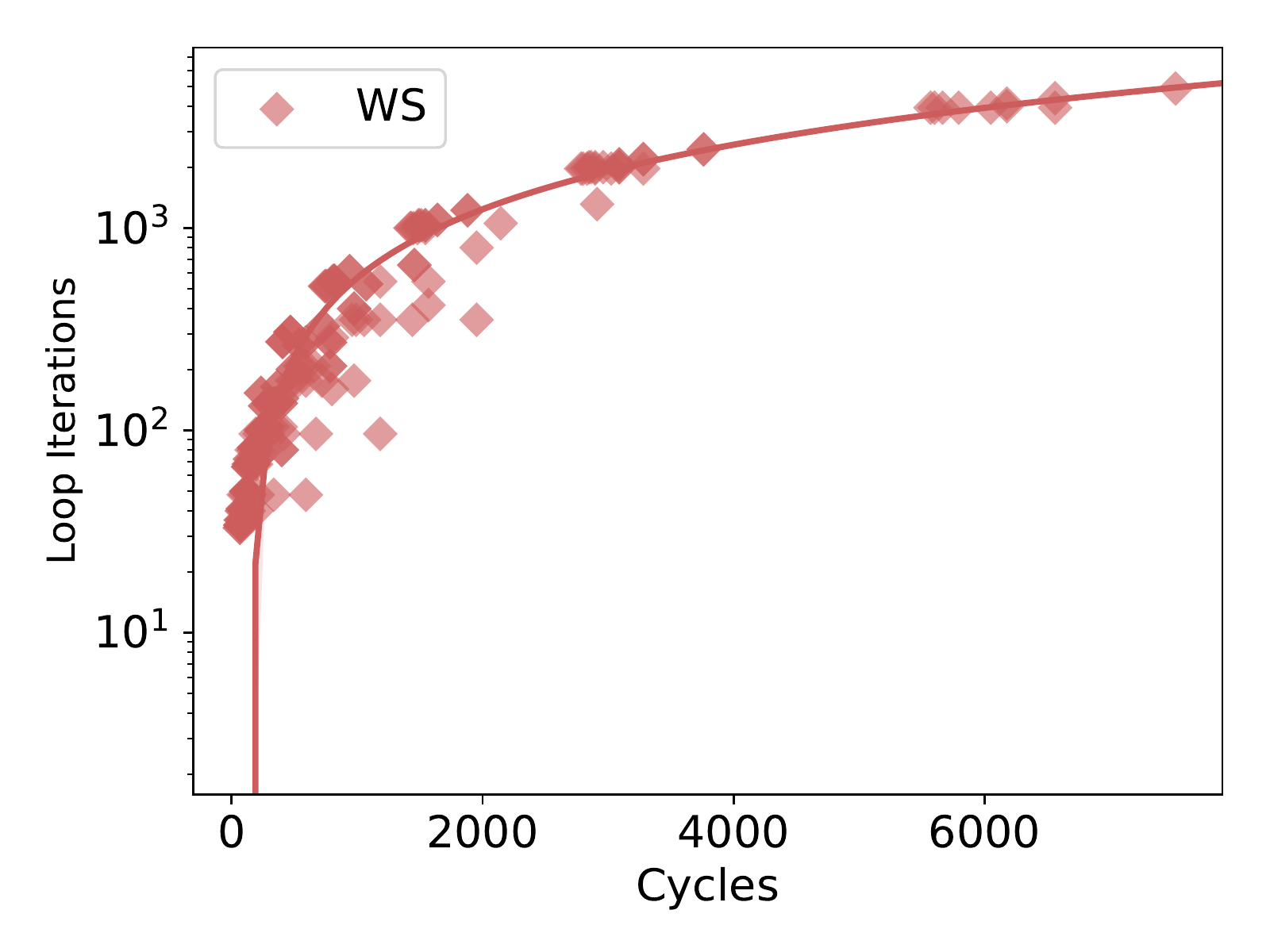}  
  \caption{Loop iteration for WS.}
 \label{fig:cycles_ratio_ws}
\end{subfigure}
\begin{subfigure}[t]{.19\textwidth}
  \centering
  \includegraphics[width=\linewidth]{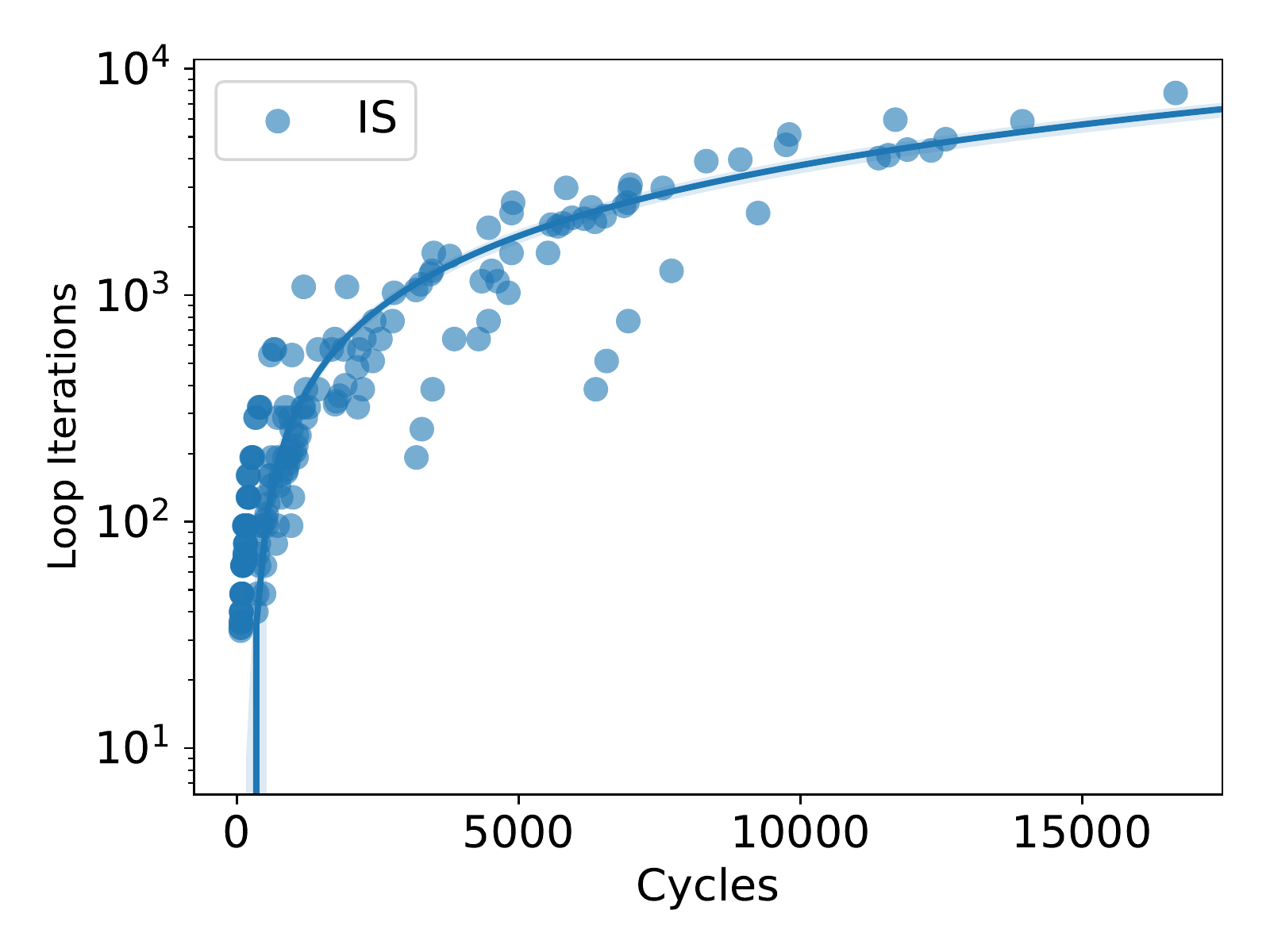}  
  \caption{Loop iteration for IS.}
 \label{fig:cycles_ratio_is}
\end{subfigure}
\begin{subfigure}[t]{.19\textwidth}
  \centering
  \includegraphics[width=\linewidth]{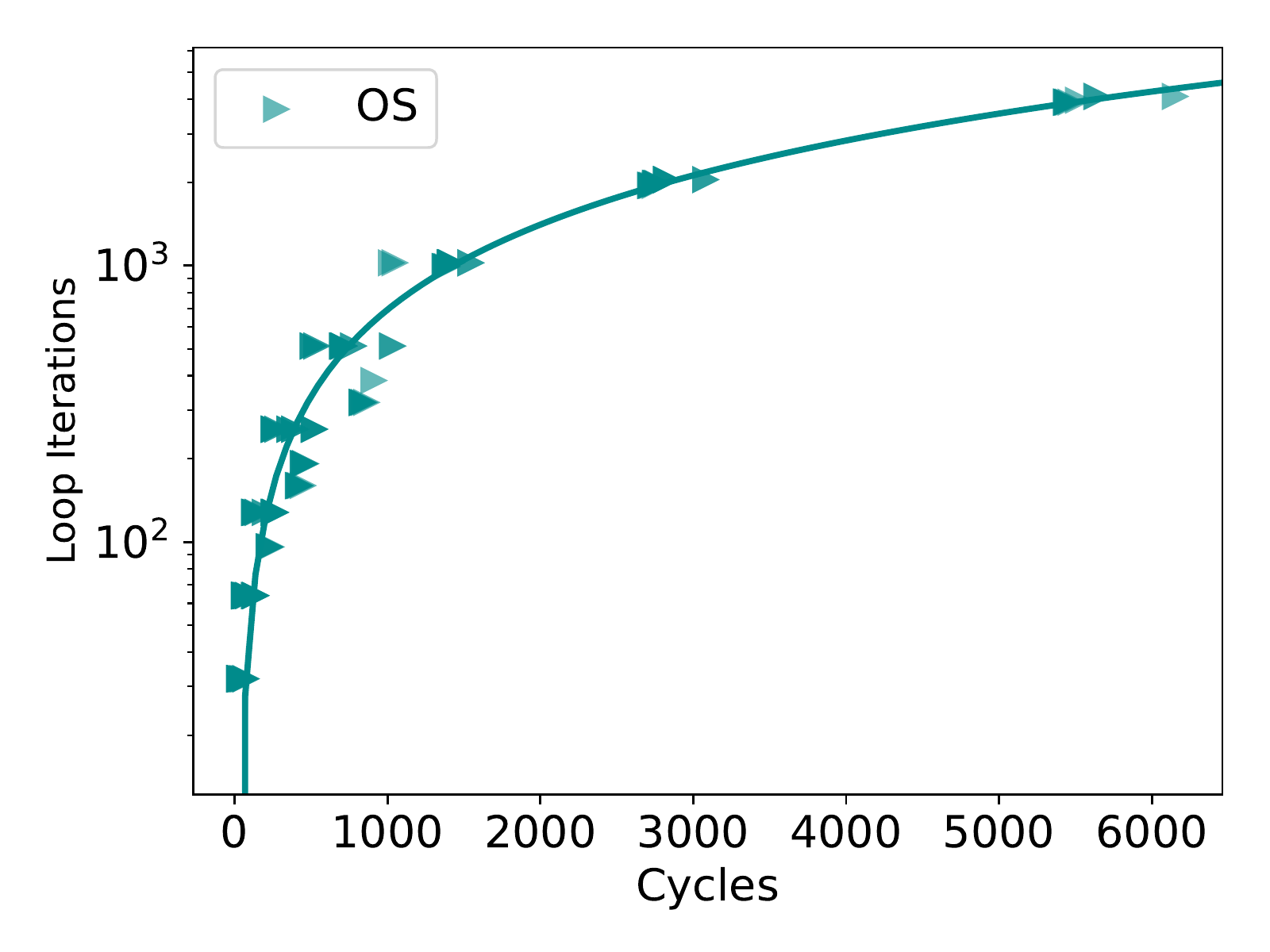}  
  \caption{Loop iteration for OS.}
 \label{fig:cycles_ratio_os}
\end{subfigure}
% \caption{Figure depicting the relation between cycles and array sizes for three dataflows.}
% \label{fig:systolic-cycle-insights}

  \caption{Given various convolution and array configuration, plotting different parameters versus cycles for three dataflows.}
 \label{fig:scalability}
% \begin{subfigure}[t]{.32\textwidth}
%   \centering
%   \includegraphics[width=\linewidth]{fig/optimal_cycles_bw_a.png}  
%   \caption{Best convolutions based on cycles. }
%  \label{fig:optimal_rate0}
% \end{subfigure}
% \begin{subfigure}[t]{.32\textwidth}
%   \centering
%   \includegraphics[width=\linewidth]{fig/optimal_cycles_bw_abc.png}  
%   \caption{Best convolutions based on cycles, read and  write bandwidth.}
%  \label{fig:optimal_rate1}
% \end{subfigure}
% \begin{subfigure}[t]{.32\textwidth}
%   \centering
%   \includegraphics[width=\linewidth]{fig/optimal_cycles_bw_bc.png}  
%   \caption{Best convolutions based on read and write bandwidth.}
%  \label{fig:optimal_rate2}
% \end{subfigure}
% \caption{Plotting the optimal array size and dataflow for each convolution based on different metrics.}
% \label{fig:systolic-optimal}

%\end{minipage}
\end{figure*}

For hardware designers, information on different dataflow performance patterns is essential when designing new mapping strategies.
To test our simulator's generality and scalability, we measure runtime and bandwidth for 4,050 combinations of array configuration ($A_h = 2,4,8,16,32, A_w = 64/A_h$) and convolutions~($H/W = 2,4,8,16,32$, $F_h/F_w/C= 1, 2, 4$, $N=1,2,4,8,16,32$) on the three dataflows.
% we again model the performance of various convolution parameters, dataflow and PE array placement and we list all the configurations in Table~\ref{table:exp-parameter}.
%Fig.~\ref{fig:scalability} plots cycles versus read bandwidth, execution time, and loop counts on the three dataflows.

\noindent \textbf{Simulator scalability.}
Fig.~\ref{fig:cycle-exec-time} plots the simulator execution time versus the cycle counts for each simulation.
The execution time is roughly proportional to the cycle count, since our simulator faithfully reflects behavior of each processor.
At most, our simulator may require over 10 minutes for simulation.
\xxx[as]{This seems to contradict ``at most 7.2 seconds'' from subsection C.}
Future work could reduce execution time by
building a lookup table to skip duplicated behavior or by adding parallelism to the simulation engine.

\noindent \textbf{Dataflows.}
Fig.~\ref{fig:cycle-bw-read} plots the SRAM read bandwidth at peak (the maximum bandwidth times the duration) versus cycle time.
OS has the highest read bandwidth overhead while WS requires the least. Though Fig.~\ref{fig:cycle-exec-time} highlights OS can achieve the shortest runtime in cycles, designers can choose the dataflow according to hardware requirements.

\noindent \textbf{Array configuration.}
Our simulator can help designers observe general ``rules'' about performance.
\Cref{fig:cycles_ratio_ws,fig:cycles_ratio_is,fig:cycles_ratio_os} plot the relationship between cycles and array structures.
The loop iteration count is proportional to cycle count.
We can calculate loop iterations as $\lceil D1/A_h\rceil \times \lceil D_2/A_w\rceil$, where $D_1 = F_H\cdot F_w\cdot C, D2=N$ for WS, $D_1 = F_H\cdot F_w\cdot C, D_2=E_h*E_w$ for IS and $D1 = N,D_2= F_H\cdot F_w\cdot C$ for OS.
With this general rule, we can always get the minimal execution time by choosing the array structure that minimizes loop iterations.

%For example, WS requires weights to be stationary for $E_h$ and $E_w$ loops, while the other four dimensions are $F_h$, $F_w$, $C$ and $N$, where the first three get mapped to array rows $A_h$ and the last get mapped to array columns $A_w$.
%when a convolution is fixed,
% \xxx[as]{I admit I got a little lost by this stufff... perhaps just because it's so dense. I think we also need to be careful here: is this data that we can reasonably say that reviewers directly asked for? I think we can only include it if we can tie it to some question from the reviews.}

\noindent\textbf{Benefits.}
The evaluation on 4050 data points shows that our simulator scales to various convolutions. Algorithm designers can use it to choose the best dataflows and array configuration for a convolution.

\section{Case Study: ACAP AI Engine}
\label{sec:aie}

A common approach to hardware--software co-design is to start simple and, guided by bottlenecks, build up a more sophisticated architecture.
% In practice, it is friendly for hardware designers to start simple and gradually build up complexity to their designs. %Operational-wise tracing and reusable representation of our simulation flow facilitate.
This section uses the
\equeue\ to simulate a real-world architecture: Xilinx's AI Engine in Versal ACAP~\cite{swarbrick2019network}.
We show our simulation result matches the AI Engine simulator~\cite{ai-engine-fir}, while the high-level simulator allows architectures ignore real-word constraints like bandwidth and gradually introduce them with low programming cost. During this process, the EQueue visualized tracing can guide designers to improve their designs.
%-- a real-world architecture - Xilinx AI Engine in Versal ACAP \cite{swarbrick2019network} -- to give an example on how \equeue\ helps lower programming overhead and operational-wise tracing facilitate designers to improve their designs. 
%from one sequential processor to multiple parallel processors, from unlimited resources to realistic bandwidth constraints and from wasting computation power to just-enough computation units with minimized warm-up stage. 
% This case study demonstrates how our simulation truthfully reflects the changes in hardware so that guided by simulation results, one can improve the architecture step by step: from one sequential processor to parallel processors, from unlimited resources to realistic bandwidth constraints and from wasting computation power to just-enough computation units with minimized warm-up stage. This kind of gradual modifications on the designs is only possible when we separate hardware representation from simulation and only when the representation is highly extensible as \equeue.

%We demonstrate how to start with a simple design and then add more complexity by modifying hardware mapping strategy through \equeue\ without re-engineering the architecture, which is laborious with existing simulator like gem5~\cite{binkert2011gem5}.
%We also show how operation-wise visualizable tracing provided by the EQueue simulator can guide designers to better designs.

\subsection{Versal ACAP}

Xilinx's Versal adaptive compute acceleration platform (ACAP) is a reconfigurable platform that includes programmable logic, ARM cores, and \emph{AI Engines}, which are specialized vector units~\cite{gaide2019xilinx, versal-acap, swarbrick2019network}. %It separates data movement architecture from compute architecture and hardens network-on-chip (NoC) to connects compute architectures, providing powerful heterogeneous accelerator for machine-learning applications. 
The AI Engine
is a fixed array of interconnected VLIW SIMD processors optimized for signal processing and machine learning.
% There are various mapping strategies that can be applied to the fixed AI engine array.
%As each AI engine core has fixed architecture, the design complexity lies on mapping strategies which can be hard to get an one-time perfect design and check correctness without detailed tracing.
%By running FIR on an AI engine array, we demonstrate our simulation flow can guide progressive changes on hardware to better utilize real-world architectures. 
%AXI-Stream interconnect blocks. %It can deliver comparable performance to FGPA on compute-intensive applications but 8x compute capacity and 50\% power consumption~\cite{ai-engine}. Figure\comment{ai-engine, shall I put one?} illustrates the AI engine array composition.
%As an solution to the decreasing benefits from technology scaling and an evolution from traditional field-programmable gate array (FPGA), ACAP raises the abstraction level by supporting flexible custom computation and data movement while fixing optimized architecture for the ease of use~\cite{gaide2019xilinx}. 

%Our \equeue\ can provide accurate performance estimation without actually programming and running the AI engines. Also, detailed visualization of each operation facilitates the architecture designer to analyze and better utilize hardware and design compact architecture.
\subsection{FIR}

%An analog signal can be converted to digital signal with two operations: slicing and quantization. 
A finite impulse response (FIR) filter is a common signal processing primitive that responds to inputs of finite duration.
An FIR operation filters and accumulates on a sliding-window.
Given a series of discrete input samples $x$ and $N$ coefficients $c$, the output samples $y$ are calculated as:
\begin{align*}
    y_n = \sum^{N-1}_{k=0} c_k\cdot x_{n+k}
\end{align*}
Xilinx's AI Engine programming tutorial~\cite{ai-engine-fir} uses a FIR filter as an example to demonstrate the hardware's flexibility and capabilities.
In this case study, we implement the same FIR example using the \equeue\ dialect to demonstrate how the language and simulation engine can easily model an existing programmable architecture.
We compare our simulator's reports to those from Xilinx's own, hand-written, closed-source simulator to ground the results.

% In this case study, we implement the same FIR example as the Xilinx AI Engine tutorial~\cite{ai-engine-fir} using EQueue simulation to demonstrate how it guides designers to improve real-world architectures.
%\xxx[as]{Can we clarify that this is an ``official'' Xilinx tutorial? As discussed with Steve, this basis on a ``real'' AI Engine use case seems to strengthen our argument that this is a realistic scenario.}
The Xilinx FIR tutorial uses a filter with 32 complex, asymmetric coefficients and a digital series of length 512.
Each value occupies 32 bits.
%We demonstrate how the EQueue simulation flow can guide designers to improve real-world architectures.

%Fig.~\ref{fig:fir} shows a FIR filter of  accumulating on a input sample series.
% \begin{figure}[h]
%  \centering
%  \includegraphics[width=0.45\textwidth]{fig/FIR_Filter.jpg}
%  \caption{Finite Impuse Response}
%  \label{fig:fir}
% \end{figure}

%In this case study, we study an FIR filter of length 32 to an input digital signal of 512 samples. 
%We would like to find the best mapping strategy of FIR operation on AI engines to best utilize the resource sand maximize the throughput. 

\subsection{Case 1: Unlimited Resources}
\label{sec:aie-case1}

We start with a basic 1-processor implementation and use empirical measurements to improve the design.
%
% As a hardware designer, the ideology of designing architecture is always to start simple and add complexity.
% % during which one can gradually gain with deeper understanding on the architecture.
% Therefore, we start with a single processor. %and assume there is no constraint on bandwidth. %By modifying the EQueue-program, we will gradually learn how to maximize throughput before and after introducing bandwidth constraints on AI engine.
%
We can use the AI Engine's intrinsics: \code{mul4} and \code{mac4}.
% They are VLIWs that has corresponding C++ API.
On each cycle, \code{mul4} computes on 4 parallel lanes to perform 8 multiplications where each lane performs 2~\cite{versal-acap-user-guide}.
\code{mac4} works in the same way. 
Analytically, therefore, it should take 16 cycles to compute 4 outputs for a filter length of 32.
%For every 16 instructions, the filter register remains the stable, while data register should be updated with 4 inputs.

%Though \code{mul4} and \code{mac4} have no corresponding operation in MLIR existing dialects, 
We follow~\cref{sec:language-external} to self-define \code{mul4} with \code|equeue.op("mul4", {ofmap, ifmap, filter})|.
In the simulator library, an operation with the ``mul4" signature reads from a buffer, computes 4 lanes with 2 computation at each lane per cycle, and writes to the buffer.
We define the \code{mac4} operation the same way.
This pseudocode shows the MLIR generator
for a single-core implementation, where \code{ifmap}, \code{ofmap} and \code{filter} are buffers:
\begin{lstlisting}
start = equeue.control_start()
equeue.launch(...) in (start, ai_engine){ 
    equeue.op("mul4", {ofmap, ifmap, filter})
    for 0 to 11:
        equeue.op("mac4", {ofmap, ifmap, filter})
    ifmap_tensor = equeue.read(sin)
    equeue.write(ifmap_tensor, ifmap)
    for 0 to 4:
        equeue.op("mac4", {ofmap, ifmap, filter})
    ofmap_tensor = equeue.read(ofmap);
    equeue.write(ofmap_tensor, sout)
}
\end{lstlisting}
Our EQueue simulation reports
2048 cycles to generate 512 outputs, close to Xilinx AI Engine simulator's result of 2276 cycles~\cite{ai-engine-fir}.
The Xilinx simulator also models other factors in performance,
including loop control costs, synchronization overhead, etc.
The EQueue simulation engine's throughput is slightly higher because it does not model these overheads.
%Therefore AI engine simulator report has lower throughput than the EQueue simulation engine.
%Assuming the AI engine array is running at 250~Ms, the throughput is 250~Gsps. In general, the throughput matches the one reported by AI engine simulator by~\cite{ai-engine-fir}, which is 225.28 Msps, since low-level hardware has synchronization overhead our simulator not consider.
%512\times32/8/(2048\times10^{-9})
% We define the stage where not all runnable AI engines are start to run as \emph{warm-up stage}. In this example, the warm-up stage is 0 since there is only 1 runnable AI engine it immediately starts. 

% Our EQueue simulator also outputs a tracing file in event trace format that can be visualized in \url{chrome://tracing}.
% Fig.~\ref{fig:single-kernel-fir} accurately reflects the operation-wise tracing of FIR implemented with a single processor, where the green slot is \code{mul} operation, red slots are \code{mac} operation, the blue slot indicates installing and the x-asis is cycle counts with $1 \mu s$ as one cycle. We omit memory tracing as they are less helpful in this example.
% %is fully occupied by computation as the bandwidth is unlimited, i.e. the processor got all data on first cycle.
% \begin{figure}[h]
%  \centering
%  \includegraphics[width=0.45\textwidth]{fig/singleKernel.png}
%  \caption{Visualizing operation-wise tracing of FIR implemented with single processor.}
%  \label{fig:single-kernel-fir}
% \end{figure}

\subsection{Case 2: Optimizing Case 1}
\label{sec:aie-case2}

The next step for a hardware designer is to incrementally increase the design's complexity to attain higher throughput.
In an ideal world, since \code{mul4/mac4} computes 4 lanes, each with 2 operation per cycle, we could pipeline $32/2=16$ processors to maximize throughput.
Due to bandwidth constraints, Xilinx's FIR tutorial simulates 4 processors rather than 16.
Using the EQueue model, we can first model the full 16-processor pipelined system and then introduce more realistic constraints to measure their effect on performance.
% The \equeue\ is high-level abstracted and not limited to real-world constraints. We can therefore pipeline 16 processors and then introduce real-world constraints to check on the performance.

The modification to our EQueue program is straightforward.
Instead of one processor executing 16 sequential operations, we now create 16 processors, where each processor completes one \code{mul4}/\code{mac4} operation.
We show the simplified control flow:
\begin{lstlisting}
start = equeue.control_start()
for k in 0..16:
    equeue.launch(...) in (start, ai_engine[k]){ 
        ifmap_tensor = equeue.read(sin)
        equeue.write(ifmap_tensor, ifmap)
        equeue.op("mac4", {ofmap, ifmap, filter})
        ofmap_tensor = equeue.read(ofmap);
        equeue.write(ofmap_tensor, sout)
    }
equeue.await()
\end{lstlisting}
The simulation engine reports 143 cycles to produce outputs for 512 inputs.
This matches the expected performance because
pipelining 16 processors requires 15 cycles to warm up.
%The throughput is
%512\times32/8/(143\times10^{-9})=
%3.6~GBps. The steady-state bandwidth should be 4~GBps as pipelining 16 processors requires 15 cycles to warm up.

%, though we expect the steady-state bandwidth to be 16 GBps. %Fig.~\ref{fig:16-kernel-fir} depicts the 16 parallel processors and explains the reason: pipeling 16 processors requires 15 cycles to warm up till last AI engine core gets the input.

% \begin{figure}[h]
%  \centering
%  \includegraphics[width=0.45\textwidth]{fig/16Kernel.png}
%  \caption{Visualizing operation-wise tracing of FIR implemented on 16 processors. Memory tracing omitted.}
%  \label{fig:16-kernel-fir}
% \end{figure}

\subsection{Case 3: Limited Bandwidth}
\label{sec:aie-case3}

% So far we have assumed bandwidth is unlimited while the only constraint is computation power. Now we would like to add the real world constraint on bandwidth and try to run the same code in the previous case. 

%It is effortless to add bandwidth constraints in EQueue-program. 
The AI Engine is constrained by the 32-bit bandwidth of its AXI4-Stream I/O interfaces~\cite{versal-acap-user-guide}.
%By default, \code{read} and \code{write} assume unlimited bandwidth. 
To add bandwidth constraints, we need only add a connection (\cref{sec:language-structure}) and update the reads and writes accordingly:
% The only modification to the code is changing bandwidth in connection creation operation.
\begin{lstlisting}
conn_in = connection("Streaming", 32);
conn_out = connection("Streaming", 32);
...
ifmap_tensor = equeue.read(sin, conn_out)
equeue.write(ofmap_tensor, sout, conn_out)
\end{lstlisting}
Adding this bandwidth constraint entails only simple, local changes to the EQueue program;
extending a custom simulator, in contrast, could require invasive modifications.
According to our simulation engine, it takes 588 cycles to generate 512 outputs,
%and the throughput is only %$512\times32/8/(588\times10^{-9})=
%870.7~Mps
including 79 cycles to warm up.

\begin{figure}[t]
 \centering
 \includegraphics[width=0.45\textwidth]{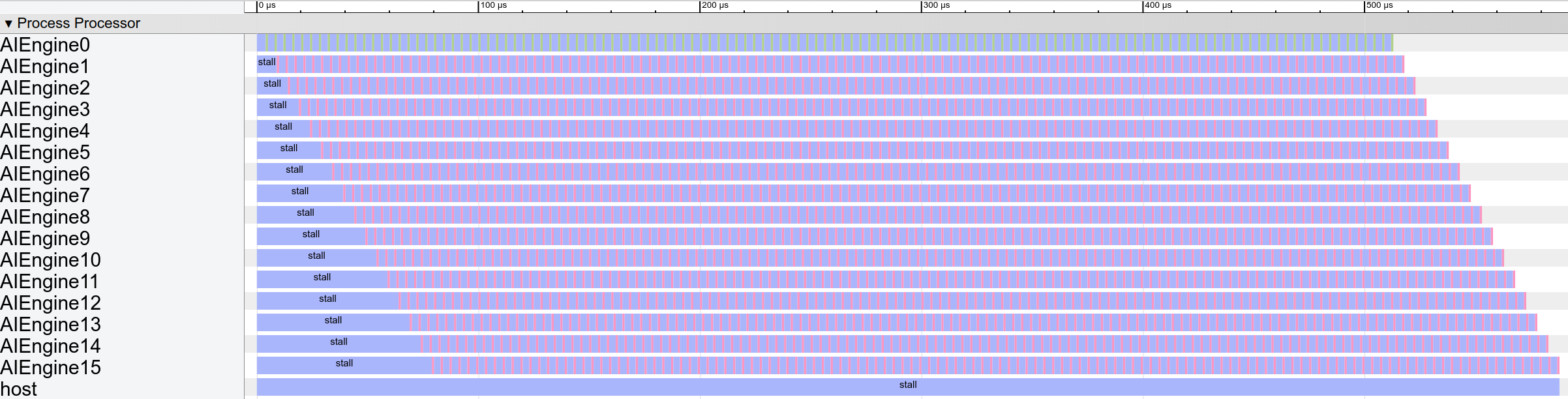}
 \caption{Visualizing operation-wise tracing of FIR implemented with 16 processors with limited bandwidth.}
 \label{fig:16-limited-fir}
\end{figure}

To understand the reason for reduced throughput, Fig.~\ref{fig:16-limited-fir} shows the operation-wise tracing with visualized via the Chrome browser, where green slots are \code{mul} operations,
red slots are \code{mac} operations, blue slots indicate installing and the $x$-axis shows cycle counts where $1 \mu s$ stands for one cycle.
% We omit memory tracing as they are less helpful in this example.
%
For every 4 cycles, each processor operation stalls for 3 cycles.
The stalls are the result of the 32-bit bandwidth constraint:
it takes 4 cycles to transmit 4 inputs, but computation only takes 1 cycle to consume these values.
For each AI Engine's attempt to start computation, it waits for its preceding core compute (1 cycle) and pass values to it (4 cycles), so the warm-up stage takes $5\times16-1=79$ cycles.

\subsection{Case 4: Optimizing Case 3}
\label{sec:aie-case4}

\begin{figure}
 \centering
 \includegraphics[width=0.45\textwidth]{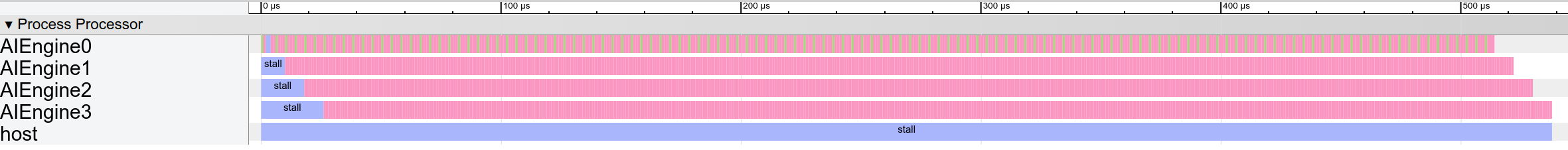}
 \caption{Visualizing operation-wise tracing of FIR implemented with 4 processors with limited bandwidth.}
 \label{fig:4-limited-fir}
\end{figure}

Our bandwidth-constrained model shows that 75\% of the hardware's computation power is wasted, i.e., we stall on 3 of every 4 cycles.
To balance the system and avoid wasting area and power, a designer can reduce the 16 processors to 4:
%
% Because when we introduce the bandwidth constraint, the computation power is wasted by 75\%, i.e., every processor stalls 3 cycles in every 4 cycles, we find that the best processor count is 4, rather than 16. 
%
% To avoid wasting computation power with busy waiting, we can modify the design:
%, where we lower 16 processors to 4, each with 4 sequential operations: 
%With MLIR generator, the core computation is express in pseudo code.
\begin{lstlisting}
start = equeue.control_start()
for k in 0..4: // 16 -> 4 cores
    // 1 -> 4 sequential computations
    equeue.launch(...) in (start, ai_engine[k]){ 
        ifmap_tensor = equeue.read(sin, connection_in[k])
        equeue.read(ifmap_tensor, ifmap)
        equeue.op("mac4", {ofmap, ifmap, filter})
        equeue.op("mac4", {ofmap, ifmap, filter})
        ofmap_tensor = equeue.read(ofmap);
        equeue.write(ofmap_tensor, sout, connection_out[k])
        equeue.op("mac4", {ofmap, ifmap, filter})
        equeue.op("mac4", {ofmap, ifmap, filter})
    }
equeue.await()
\end{lstlisting}
Our EQueue simulation engine reports that generating 512 outputs requires 538 cycles, which matches Xilinx's simulator result of 539 cycles.
%and the throughput is 
%512*32/8/(538*10^{-9})=
%3.8~GBps. 
Warm-up takes 26 cycles, which is much faster than the previous case.
Fig.~\ref{fig:4-limited-fir} visualizes the operation trace for the balanced 4-processor system: there is no stalling once the processors have warmed up.

\noindent\textbf{Benefits.}
Typical simulation tools can make it challenging for software designers to identify hardware bottlenecks.
This case study advocates an opposite approach:
designers can start with a simple design and gradually add real-world constraints to examine their effects on performance.
Our EQueue approach requires modest modification at each step, but effectively guides users to improve their design.

Our EQueue-based approach matches the results of Xilinx's existing AI Engine simulator tool.
Thanks to its high-level abstraction, the EQueue-based simulator is much faster: the 4 processor implementation takes 0.07~seconds, while the AIE simulator first requires 5 minutes for compilation and then 3 minutes for simulation.
Also, due to its focus on low-level details, the AI Engine implementation is spread across six separate files. %where each core is defined separately with different inputs.
Any updates to the interface or mapping strategy requires substantial work to implement and recompile.

\section{Related Work}

Because simulation is a critical part of a hardware design workflow, it is an old and well-studied research topic.
Space precludes a complete census of all approaches to simulation, but we discuss the most closely related techniques here.

The EQueue simulation flow takes inspiration from hardware modeling languages~\cite{li2020heterohalide, pu2017programming, hegarty2016rigel, hegarty2014darkroom, auerbach2010lime, rong2017programmatic, lai2019heterocl}.
It differs from RTL simulation with its higher-level representation and focus on an intermediate representation that can be transformed by compiler passes.
We designed the \equeue\ because existing languages and MLIR dialects cannot represent the core concepts required for flexible, high-level simulation:
fine-grained concurrency, contention for shared heterogeneous hardware resources, and data movement constraints.

% We choose to implement \equeue as existing dialects cannot fully specify hardware configuration we need. 

% \paragraph{Hardware modeling languages}
% There are many categories of programming languages for designing hardware,
% from compilers for domains like image processing~\cite{li2020heterohalide, pu2017programming, hegarty2016rigel, hegarty2014darkroom} or machine learning~\cite{sharma2016high, george2014hardware},
% languages for general accelerator design~\cite{auerbach2010lime, rong2017programmatic, lai2019heterocl},
% and IRs targeting software~\cite{kingma2018glow, chen2018tvm,lattner2004llvm,pilato2013bambu} and hardware~\cite{wang2019lnast, izraelevitz2017reusability, nigam2021compiler, nigam2020predictable, schuiki2020llhd}. %A good hardware description language articulate an architecture by itself.
% These approaches inspired \equeue\ dialect's separation of the hardware representation from the simulation engine.
% We take the advantage of MLIR~\cite{lattner2020mlir} to save effort on the IR's scaffolding and to interoperate with
% existing dialects.
% %cross-boundary compilation, common AST among different dialects and easy extensibility. 
% %all share the same AST with different abstraction level. It is designed for easy extension and supports high-level language features include but not limited to value dominance inside regions and library-like dialects.

\paragraph{RTL simulators}
Most RTL development tools have accompanying simulators~\cite{intel-simulator,hls-simulator,vivado-simulator, kim2016strober, kim2017evaluation}.
%Strober~\cite{kim2016strober} provides cycle accurate estimation for FPGA.
%MIDAS~\cite{kim2017evaluation} generates FPGA-accelerated simulators from RISC-V RTL.
RTL simulation can faithfully model a complete hardware design, but implementing a design in RTL requires specialized hardware expertise and carries a high engineering burden.
An alternative is integrating a more abstract simulation with RTL using a multi-level tool such as PyMTL~\cite{lockhart2014pymtl}.
We view the \equeue\ as a complement to these frameworks that makes it easier to generate and transform higher-level models before completing a more detailed implementation.

%Charm, a domain specific language supporting Closed-form High-level Architecture Modeling. Charm enables mathematical representations of mutually dependent architectural relationships to be specified, composed, checked, evaluated and reused. The language is interpreted through a combination of symbolic evaluation (e.g., restructuring) and compiler techniques (e.g., memoization and invariant hoisting), generating executable evaluation functions and optimized analysis procedures.

%PAAS: A system level simulator for heterogeneous computing architectures ~\cite{liang2017paas} a system level simulator for heterogeneous CPUs-accelerator systems which integrates a modified version of gem5 and the HDL simulator Verilator to enable accurate simulation of dynamic interactions

\paragraph{Application-specific simulators}

% We can find in the
% literature other proposals which focus on modeling specific
% programs or program classes (patterns). For example, Meng
% and Skadron propose in [25] a framework that automatically selects the best parameters for an Iterative Stencil
% Loops (ISL) application, given some parameters of the
% domain and some features of the target GPU. ISL is a technique applied in many domains, including molecular
% dynamics simulation and image processing, that distribute
% computations into overlapping regions (tiles), with neighboring regions interacting via halo zones. Similarly, Choi
% et al. present in [26] an auto-tuned implementation of the
% Sparse Matrix-Vector (SpMV) multiplication. The difficulty
% in computing with sparse matrices is related to using compression algorithms (such as BCSR or ELLPACK [27]) that
% represent matrix data as lists. A

% this~\cite{choi2019cnn} propose a simulator for designing CNN inference accelerator using bit optimization and tiling methods. 
% potential problems: too specific to certain data flow.
% Scale-Sim provides a look into the simulation for systolic array-based accelerators~\cite{samajdar2018scale}. (specific to systolic arrays)

Many efforts have constructed architecture-specific simulators,
for domains including sparse linear algebra~\cite{choi2010model},
stencils~\cite{datta2009optimization},
and DNN inference~\cite{choi2019cnn,samajdar2018scale,kwon2019understanding,parashar2019timeloop}.
While these simulators are fast and accurate,
they are challenging to construct from scratch.
The \equeue\ dialect provides a faster way to build simulators.

%The simulator could easily become out of the scope when users switch to a new workflow or backend. 
%The EQueue generic simulation engine adapts to all EQueue inputs and the effort to modify an input design is minimized.% with reusable compiler passes. 
%supports simple designs changes without hindering much of the ability to simulate efficiently and accurately.
% Our dialect avoids this shortcoming by a clear separation between representation and simulation. It allows our simulator to support many designs without hindering much of the ability to simulate efficiently and accurately. The lowering structure in MLIR further makes our dialect easily extensible and reusable.

% \paragraph{General-purpose simulators}
% There are many general purpose models targeting specific backends.
% Aladdin~\cite{shao2014aladdin} simulates C programs by building dynamic data dependence graphs. % without generating RTL.
% Extensions to gem5~\cite{binkert2011gem5} such as gem5-gpu~\cite{power2014gem5gpu} and gem5-Aladdin~\cite{shao2016co} model heterogeneous systems and the interactions between components in an SoC.
% % gem5-gpu~\cite{power2014gem5gpu} simulates heterogeneous systems based on gem5 CPU simulator~\cite{binkert2011gem5}. %targeting CPU simulation.
% % Gem5-Aladdin~\cite{shao2016co} captures dynamic interactions between accelerators and the systems on chip platforms.
% These workflows do not focus on representing arbitrary, fine-grained concurrency and data movement.
% They also do not expose intermediate forms of the hardware design to compiler passes for analysis and transformation.

\paragraph{MLIR methodology}

It is appealing to use existing MLIR dialects that already offers various transformations and ways to express computations and hardware.
For example, MLIR's \emph{Async} dialect~\cite{async-dialect} models asynchronous execution.
It cannot, however, associate code with specific processing units in a hardware structure.
CIRCT~\cite{eldridgemlir} is an ongoing project to apply MLIR's methodology to hardware design tools. It encompasses many dialects, including a \emph{Handshake} dialect representing asynchronous processes
%First-in First-out (FIFO) communication channels
that can compile
to a \emph{FIRRTL} dialect for circuit-level transformations and then to \emph{LLHD} dialect to describe RTL.
The \emph{HIR} dialect~\cite{majumder2021hir} describes hardware with explicit scheduling and binding, which serves as a better IR than pure LLVM for HLS-like compilation from software to hardware.
Both HIR and the CIRCT dialects are abstractions for generating concrete hardware implementations, not high-level abstractions for modeling concurrency and data movement for efficient simulation.
% FPGA, not considering expressiveness for multiple concurrent processors. They can serve as one of the backends for \equeue.
The \equeue\ differs by explicitly representing execution units and mapping event-triggered computations onto them.

%A new Hw dialect~\cite{hw-dialect} represents hardware at an RTL-like abstraction level, which is more detailed than EQueue's high-level model.
%a event trigerred arbitrary hardware configuration with explicit data movement.

% I don't think the GPU dialect is quite relevant enough to include. --AS
% On the other hand, GPU dialect models GPU execution~\cite{gpu-dialect}. However, it is an abstraction for GPUs and the architecture is fixed. GPU dialect can be a lowering target for the \equeue.

%the programing of the accelerator: either use software developed by the manufacturer, or standardized API. "It is important to stress that the use of standard APIs may provide code portability between different platforms, but this does not translate into performance portability: currently, device specific, performance-oriented fine-tuning of an application is required in order to fully exploit the capabilities of an accelerator." accelerators, CPUs and memory system. 

\paragraph{Compiler-driven DSE}
Compilation is an efficient way to perform design space exploration (DSE), especially in the specific domain of tensor computations.
Interstellar~\cite{yang2020interstellar} uses Halide~\cite{ragan2013halide} to explore DNN accelerator designs.
Union~\cite{jeong2021union} uses MLIR programs as inputs to optimize spatial DNN accelerators by analyzing tensor operations expressed in Linalg or Affine dialect with MAESTRO~\cite{kwon2019understanding} and Timeloop~\cite{parashar2019timeloop} as cost models.
Similar to the EQueue methodology, these frameworks benefit from separating modeling from representation for rapid iteration.
However, all of these approaches target a specific category of computation and hardware:
they map high-level DNN dataflow mappings to synchronous PE arrays of regular structures.
The \equeue\ aims to address \emph{general} hardware simulation, including programmable architectures that do not resemble systolic arrays, such as the AI Engine (Section~\ref{sec:aie}).
EQueue may also be a good fit for extending Union with support for explicit representations of hardware components.

\section{Conclusion}
Hardware simulation frameworks need a separation of simulation from representation.
The simulation flow for EQueue programs lowers the bar for designers with abstract representations, exposes intermediate optimization stages, and makes it easy to apply changes with reusable compiler passes.
%A simulatable representation across abstraction levels lowers the bar for optimizations, analysis and transformations.

%\section*{Acknowledgements}
%This document is derived from previous conferences, in particular ISCA 2020.

%%%%%%% -- PAPER CONTENT ENDS -- %%%%%%%%

%%%%%%%%% -- BIB STYLE AND FILE -- %%%%%%%%
\bibliographystyle{IEEEtranS}
\bibliography{refs}
%%%%%%%%%%%%%%%%%%%%%%%%%%%%%%%%%%%%

\end{document}